\documentclass[twocolumn,preprintnumbers,amsmath,amssymb,
reprint,groupedaddress,amsmath,citeautoscript,flushbottom,
floatfix,superscriptaddress]{revtex4}
\usepackage{graphicx}
\usepackage{dcolumn}
\usepackage{bm}
\usepackage{amsmath}
\usepackage{amssymb}
\usepackage{amsmath}
\usepackage{booktabs}
\usepackage{multirow}
\usepackage{array}
\usepackage{color}
\usepackage{xcolor}
\usepackage[normalem]{ulem}
\usepackage{float}
\usepackage[colorlinks=true,linkcolor=blue,anchorcolor=red,citecolor=blue, urlcolor=blue]{hyperref}
\usepackage{bbm} 
\usepackage{multirow, makecell} 
\usepackage{extarrows}

\makeatletter
\def\@hangfrom@section#1#2#3{\@hangfrom{#1#2#3}}
\makeatother
\newcommand{\vc}[1]{\boldsymbol{#1}}

\begin{document}

\title{Correlation Induced Magnetic Topological Phases in Mixed-Valence Compound SmB$_6$}

\author{Huimei Liu}
\affiliation{Institute for Theoretical Solid State Physics and W\"{u}rzburg-Dresden Cluster of Excellence ct.qmat, IFW Dresden, Helmholtzstr. 20, 01069 Dresden, Germany}
\affiliation{Max Planck Institute for Solid State Research, Heisenbergstrasse 1, D-70569 Stuttgart, Germany}

\author{Moritz M. Hirschmann}
\affiliation{Max Planck Institute for Solid State Research, Heisenbergstrasse 1, D-70569 Stuttgart, Germany}

\author{George A. Sawatzky}
\affiliation{Department of Physics and Astronomy, University of British Columbia, Vancouver B.C. V6T 1Z1, Canada}
\affiliation{Stewart Blusson Quantum Matter Institute, University of British Columbia, Vancouver B.C. V6T 1Z4, Canada}

\author{Giniyat Khaliullin}
\affiliation{Max Planck Institute for Solid State Research, Heisenbergstrasse 1, D-70569 Stuttgart, Germany}

\author{Andreas P. Schnyder}
\affiliation{Max Planck Institute for Solid State Research, Heisenbergstrasse 1, D-70569 Stuttgart, Germany}

\begin{abstract}
SmB$_6$ is a mixed-valence compound with flat $f$-electron bands that have a propensity to magnetism. Here, using a realistic $\Gamma_8$ quartet model, we investigate the dynamical spin susceptibility and describe the in-gap collective  mode observed in neutron scattering experiments. We show that as the Sm valence increases with pressure, the magnetic correlations enhance and SmB$_6$ undergoes a first-order phase transition into a metallic antiferromagnetic state, whose symmetry depends on the model parameters. The magnetic orderings give rise to distinct band topologies: while the A-type order leads to an overlap between valence and conduction bands in the form of Dirac nodal lines, the G-type order has a negative indirect gap with weak $\mathbb{Z}_2$ indices. We also consider the spin polarized phase under a strong magnetic field, and find that it exhibits Weyl points as well as nodal lines close to the Fermi level. The magnetic phases show markedly different surface states and tunable bulk transport properties, with important implications for experiments. Our theory predicts that a magnetic order can be stabilized also by lifting the $\Gamma_8$ cubic symmetry, thus explaining the surface magnetism reported in SmB$_6$.
\end{abstract}

\date{\today}

\maketitle


In the recent years, the fields of band topology~\cite{Chi16} and heavy fermions~\cite{steglich_nat_review} have intertwined to form the new research direction of  topological heavy-fermion materials~\cite{Dze10,Baruselli2014,severing_andreas_Sci_Rep_15,po_yao_moebius_kondo,Kimura2018,allen_SmB6_review,nagai_liang_fu_PRL_20,Kle20,silke_paschen_Weyl_kondo_PNAS_21,broholm_weyl_mediated_magnetism_Nat_mat_21,fisk_magneto_transport}. These materials exhibit topologically nontrivial band structures together with a number of strong-correlation effects, e.g. non-Fermi liquid behavior, unconventional quantum criticality, and Kondo lattice physics. The interplay between topology and electron correlations creates various novel phenomena, whose robust nature makes them amenable for topological quantum devices~\cite{Sme18,Gil21}. Correlation induced topological properties include exceptional points in quasiparticle spectra~\cite{nagai_liang_fu_PRL_20}, giant spontaneous Hall effects~\cite{silke_paschen_Weyl_kondo_PNAS_21}, and helical magnetism induced by Weyl electrons~\cite{broholm_weyl_mediated_magnetism_Nat_mat_21}. Yet another intriguing possibility is that the magnetism of $f$-electrons may alter the band topology of the charge carriers. This is what we study here in the context of the mixed-valence, heavy-fermion material SmB$_6$~\cite{Men69,Var76,fisk_review_21}.

\begin{figure}
\begin{center}
\includegraphics[width=8.5cm]{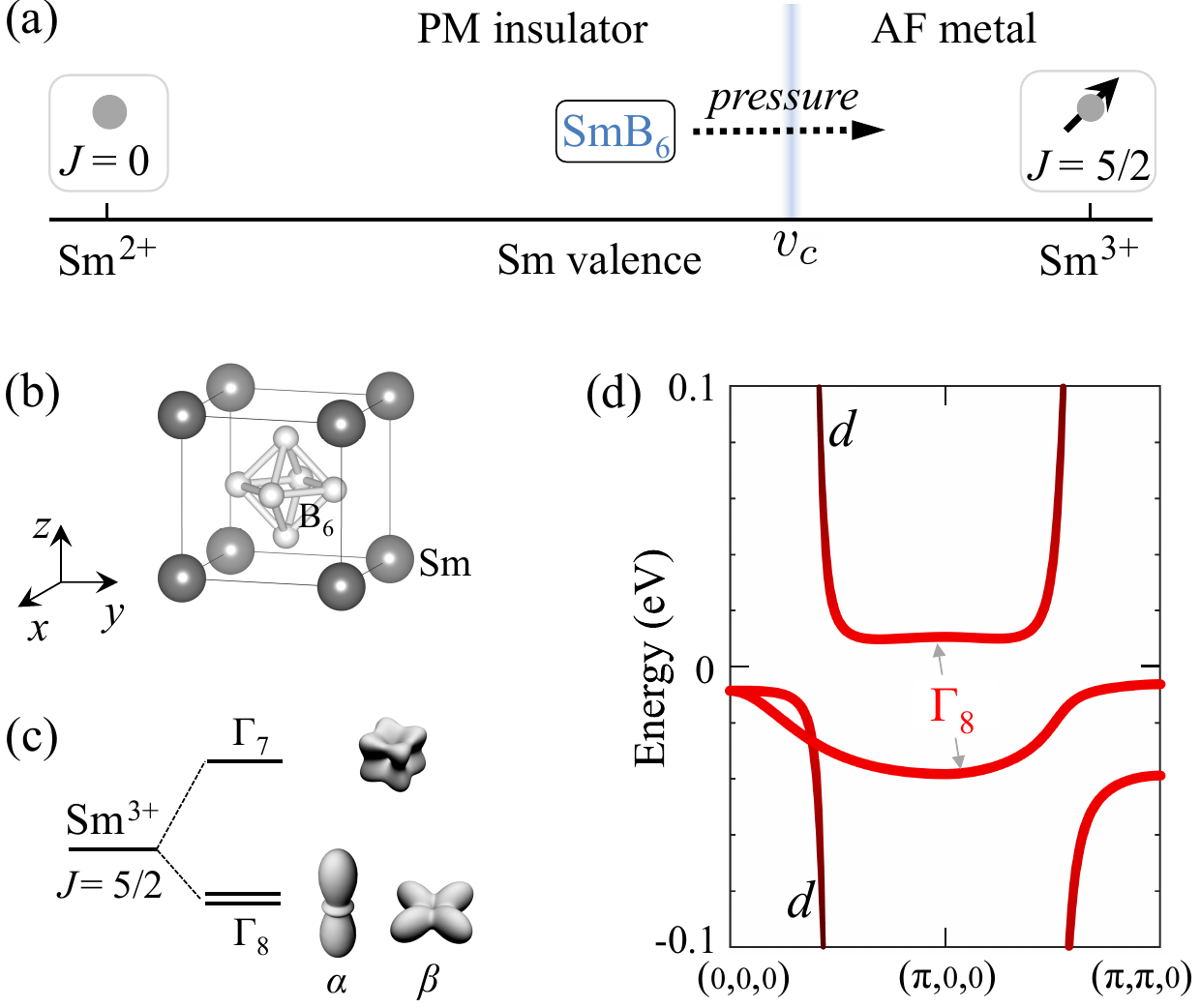}
\caption{(a) Sketch of the Sm ionic states and phase behavior of SmB$_6$ under pressure. (b) Lattice structure of SmB$_6$. (c) Crystal field splitting of the $J= 5/2$ multiplet into $\Gamma_8$ quartet and $\Gamma_7$ doublet. The $\Gamma_8$ quartet has a two-fold orbital degeneracy, $\alpha$ and $\beta$ Kramers doublets with elongated and planar spatial shapes. (d) Low energy band structure of SmB$_6$ showing an insulating gap at the Fermi level. The black and red colors indicate the contributions from the highly dispersive $5d$ and quasilocalized $4f(\Gamma_8)$ states, respectively. }
\label{fig:1}
\end{center}
\end{figure}

The electronic structure of SmB$_6$ is characterized by strongly correlated $4f$ and itinerant $5d$ electrons, whose hybridization opens up a gap at the Fermi level with nontrivial topology~\cite{Dze10,Tak11,Lu13}. The Sm valence fluctuates between Sm$^{2+}$ and Sm$^{3+}$ states. Its average value $v$ can be tuned by pressure, in favor of magnetic Sm$^{3+}$ with smaller ionic size, as sketched in Fig.~\ref{fig:1}(a). With increasing Sm valence, SmB$_6$ undergoes a discontinious insulator-to-metal transition, accompanied by antiferromagnetic (AFM) order~\cite{Bar05,Der06,Der08,But16,Emi18,Zho17}. Remarkably, this transition occurs at an intermediate valence, $v_c \sim 2.7$~\cite{But16,Emi18}, well before reaching the fully trivalent state, such that valence fluctuations coexist with magnetism. This intriguing phase transition calls for a new theoretical description. In particular, the tunable interplay between band topology and magnetism at intermediate valence may give rise to new types of topological characteristics with interesting surface properties.

In this Letter, we address the above points based on a realistic model which reproduces the insulating band structure of SmB$_6$, as well as gives rise to a low-energy collective spin excitation below the charge gap, manifesting the proximity to magnetic order. In our theory, which explicitly includes the Sm valence $v$ as a tuning parameter, the spin exciton softens and gains intensity as $v$ increases. At a critical value of $v_c$, an insulator-to-magnetic-metal phase transition of first order takes place. The transition is driven by the exchange interactions between magnetoactive Sm$^{3+}$ ions, whose density increases with pressure. Depending on the model parameters, different AFM states (such as A- or G-type orders) can form. We found that the magnetic phases exhibit band topologies that are markedly different from the insulating nonmagnetic phase. While the A-type order has nodal lines with surface states~\cite{chan_drumhead_16} at the Fermi level, the G-type order shows an indirect negative band gap with weak $\mathbb{Z}_2$ indices and corresponding Dirac surface states. In the spin polarized phase under a strong magnetic field, SmB$_6$ becomes a nodal-line Weyl semimetal. We characterize these band topologies by use of mirror Chern numbers, $\mathbb{Z}_2$ invariants, and a newly introduced ``glide-mirror-graded'' Wilson loop.

\emph{The model.}---The SmB$_6$ crystal has cubic symmetry, with B ions forming an octahedron at the cube center and Sm ions at the cube corners, see Fig.~\ref{fig:1}(b). A hybridization of the Sm $4f$ orbitals with conduction bands leads to valence fluctuations between nearly degenerate Sm$^{2+}$($f^6$) and Sm$^{3+}$($f^5$) states. The resulting  wavefunction can effectively be expressed as a coherent superposition $\sqrt{1-n} \; |f^6;d^0 \rangle + \sqrt{n} \; |f^5;d^1\rangle$, where $|d \rangle$ denotes the Sm $5d$ band states admixed with B $2p$ orbitals, and $n=v-2$ is the density of the magnetic Sm$^{3+}$ ions.

The Sm$^{2+}$($f^6$) ion has a $J=0$ singlet ground state. For Sm$^{3+}$ with the $f^5$ configuration, the lowest multiplet is the $J=5/2$ sextet, which is split by the cubic crystal field into a $\Gamma_8$ quartet and a $\Gamma_7$ doublet, see Fig.~\ref{fig:1}(c). In SmB$_6$, the $\Gamma_7$ level is about $20$~meV higher in energy~\cite{Amo19}. Moreover, the $\Gamma_7$ and $5d$ states have a negligible wavefunction overlap. Hence we neglect the $\Gamma_7$ excitation, and focus on the $\Gamma_8$ quartet, the $J=0$ singlet of Sm$^{2+}$, and the $5d$ bands to construct a low-energy Hamiltonian.

By symmetry, the transitions between $f^6(J=0)$ and $f^5(J=5/2; \Gamma_8)$ states can be described as the removal/addition of an $f$-electron with the angular momentum $j=5/2$ and $\Gamma_8$ symmetry. This mapping, which re-scales the bare $f$-electron hopping amplitudes by the coefficients of fractional parentage~\cite{Sawatzky16}, leads to the following Hamiltonian:
\begin{align}
\mathcal{H}=&E_{d}\sum_{i\sigma\gamma}d^\dagger_{i\sigma\gamma}d_{i\sigma\gamma}
+\sum_{ij}\sum_{\sigma\gamma\gamma '} t_{d,ij}
d^\dagger_{i\sigma\gamma}d_{j\sigma\gamma '}
\notag  \\
+& E_{f} \sum_{i\sigma\gamma} f^\dagger_{i\sigma\gamma}f_{i\sigma\gamma}
+\sum_{ij} \sum_{ \sigma\sigma'\gamma\gamma '}
t_{f,ij}f^\dagger_{i\sigma\gamma}f_{j\sigma'\gamma '}
\notag \\
+& \sum_{ij}\sum_{\sigma\sigma'\gamma\gamma '}
V_{ij}(d^\dagger_{i\sigma\gamma}f_{j\sigma'\gamma '}+\mathrm{H.c.})\; ,
\label{eq:H}
\end{align}
where $d^\dagger_{i\sigma\gamma}$ ($f^\dagger_{i\sigma\gamma}$) denote the creation operators of the $5d$ band ($\Gamma_8$ quartet) states on site $i$ with (pseudo)spin $\sigma$ and orbital index $\gamma$. The spin-orbital structure of the matrix elements $t_d, t_f$, and $V$ follows from the symmetries of the $5d$ and $4f(\Gamma_8)$ wavefunctions~\cite{Tak11,Tak80} (see Supplemental Material for details~\cite{supp}). The $\Gamma_8$ quartet may be either fully occupied (Sm$^{2+}$) or host a single-hole (Sm$^{3+}$). We treat this constraint on a mean-field level, and vary the energy levels $E_d$ and $E_f$ to control the $f$-hole density $n=v-2$. The hopping amplitudes, renormalized by the correlation effects, are determined by a fit to experiments~\cite{supp}.

The band structure near the Fermi level, calculated at the valence $v = 2.56$ (i.e. at ambient pressure), is shown in Fig.~\ref{fig:1}(d). We observe two flat bands originating from the $\Gamma_8$ states, and one highly dispersive band of $5d$ character. Their hybridization opens up an insulating band gap of $\sim 17$~meV, consistent with experiment~\cite{Neupane13}. The band inversion between the $\Gamma_8$ and $d$ bands at the X-point $(\pi,0,0)$ leads to a nontrivial strong $\mathbb{Z}_2$ topology with helical surface states \cite{Dze10,Tak11,Lu13}.

\emph{Spin exciton and magnetic order.}---Having the model that reproduces the low-energy electronic states of SmB$_6$, we address now its magnetic properties. In SmB$_6$, a dispersive magnetic mode is formed inside the charge gap~\cite{Fuh15}. This is a signature of strong correlations among the $J=5/2$ moments, which interact via the conduction bands or various superexchange channels (e.g., via the B orbitals). For simplicity, we model these interactions by the isotropic exchange Hamiltonian,
\begin{align}
\mathcal{H}_{\rm ex}=\sum_{\langle i j \rangle} \mathcal{J}_{ij} {\vc J}_i \! \cdot \! {\vc J}_j \;,
\label{eq:HS}
\end{align}
and calculate the dynamical magnetic susceptibility using the random phase approximation:
\begin{align} \label{chi_RPA}
\chi( {\bm q},\omega)=\frac{\chi_0( {\bm q},\omega)}{1+ \mathcal{J}_{\bm q} \chi_0( {\bm q},\omega)}\;.
\end{align}
Here, $\mathcal{J}_{\bm q}=\sum_{\bm R}\mathcal{J}_{\bm R} e^{i \bm q \cdot \bm R}$ and $\chi_0({\bm q},\omega)$ is the bare magnetic susceptibility, which can readily be evaluated using the eigenstates and energies of the electronic bands obtained above~\cite{supp}. The major contribution to $\chi_0({\bm q},\omega)$ is due to the transitions between the flat bands in Fig.~\ref{fig:1}(d), which give rise to a gapped Stoner continuum. Of our prime interest is, however, a low-energy collective mode that emerges as a sharp particle-hole bound state inside the gap~\cite{Ris00,Fuh15,Kno17}. From Eq.~\eqref{chi_RPA}, the energy $\omega_{\bm q}$ of this mode is given by the condition $\mathcal{J}_{\bm q} = -1/ \chi'_0({\bm q},\omega_{\bm q})$, and thus sensitive to the exchange parameters.

\begin{figure}
\begin{center}
\includegraphics[width=8.5cm]{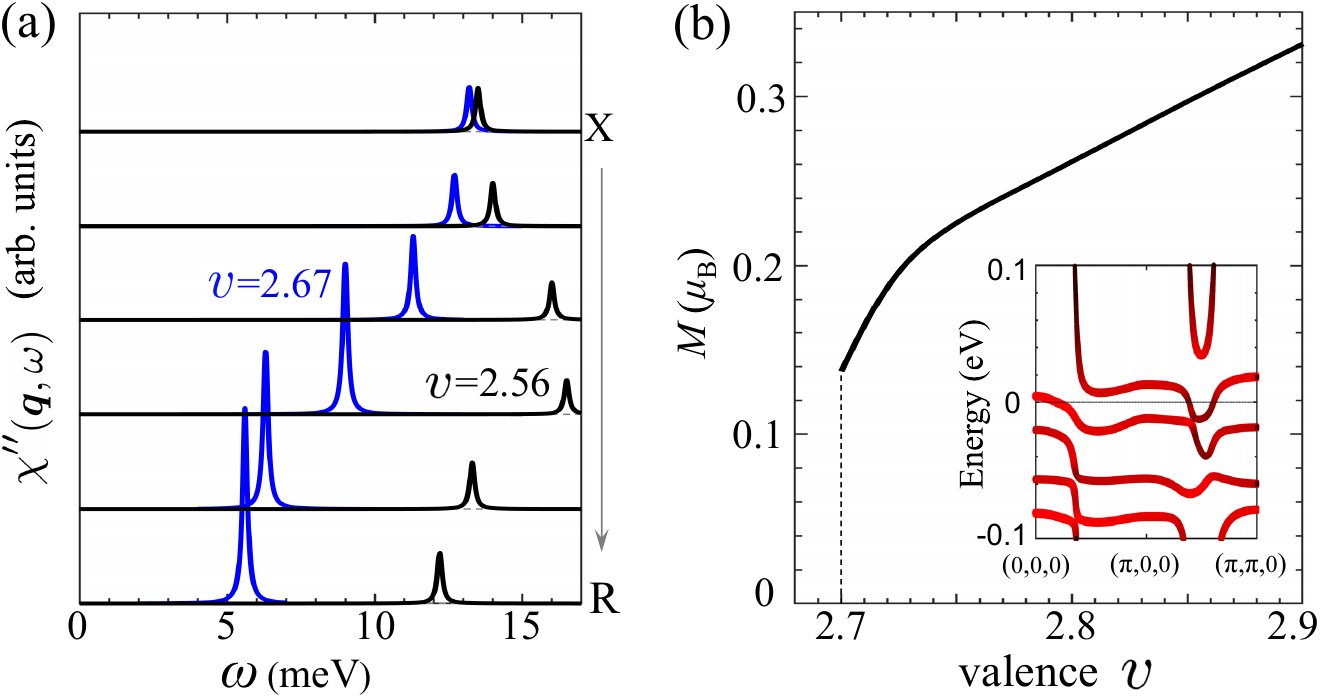}
\caption{(a) Dynamical magnetic susceptibility $\chi''({\bm q},\omega)$ for the valence $v =2.56$ (black) and $2.67$ (blue), for $\vc q$ values along the high symmetry path from $X=(\pi,0,0)$ to $R=(\pi,\pi,\pi)$. The particle-hole continuum (not shown) is above $\simeq 17$ meV. (b) The G-type ordered magnetic moment $M$ as a function of $v$. The inset shows the metallic band structure at $v=2.72$.
}
\label{fig:2}
\end{center}
\end{figure}

Having no microscopic theory for exchange interactions in SmB$_6$, we consider a minimal set of the $\mathcal{J}_{ij}$ couplings up to third-nearest neighbors. Figure~\ref{fig:2}(a) shows the spin-exciton peaks in $\chi''({\bm q},\omega)$, calculated using the parameter set $(\mathcal{J}_1, \mathcal{J}_2, \mathcal{J}_3) = (14.7, 6.5, 2.7)$~meV. The results for $v=2.56$ qualitatively agree with the inelastic neutron scattering data~\cite{Fuh15}.

We find that the spin exciton mode is highly sensitive to the valence state. It strongly softens and gains a large spectral weight as $v$ increases, see Fig.~\ref{fig:2}(a). Physically, increasing the density of magnetic Sm$^{3+}$ ions tips the balance between the exchange interactions and the $f$-electron delocalization energy, driving the system towards a magnetic instability. SmB$_6$ developes a long-range AFM order at the critical valence $v_c \sim 2.7$~ \cite{But16,Emi18}. To describe this transition, we perform a mean-field calculation for the magnetic order parameter as a function of $v$. We let the effective hopping parameters $(t_d, t_f, V)$ vary as the valence increases under pressure, and rescale them by a phenomenological factor $[1+\eta (v-v_0)]$, where $v_0=2.56$. Physically, the sign of the parameter $\eta$ is determined by two competing effects: on the one hand, the effective hoppings are reduced ($\eta<0$) by the correlations that enhance with valence $v$; on the other hand, they increase ($\eta>0$) due to the increased wavefunction overlap under pressure.  With $\eta=0.6$~\cite{note1}, a magnetic solution appears at $v_c \simeq 2.7$ via a first-order phase transition, see Fig.~\ref{fig:2}(b). Once the magnetic order is established, the spin degeneracy is lifted and the band structure gets rearranged, which induces a simultaneous insulator-to-metal transition. The feedback effects between spin order and band structure result in a discontinious transition, as observed in SmB$_6$~\cite{Bar05,Der08}.

It should be noted that the AFM ordering pattern (not yet fully identified experimentally) is decided by the choice of $\mathcal{J}_{ij}$ parameters. While the above $\mathcal{J}_{ij}$ values (motivated by the spin-exciton dispersion fits) support G-type (N$\acute{\rm e}$el) state, other structures, such as A-type order suggested by a ($\pi, 0, 0$) peak in the bare susceptibility $\chi_0 ({\bm q})$~\cite{supp}, are possible. Future experiments are required to quantify the exchange Hamiltonian in SmB$_6$.

Since the $\Gamma_8$ quartet is a spin-orbit entangled object, magnetic order also affects the orbital shape of the $f$-electron cloud and thus reduces cubic symmetry of the paramagnetic phase. Alternatively, lifting the degeneracy of the $\alpha$ and $\beta$ states of the $\Gamma_8$ quartet [Fig.~\ref{fig:1}(c)] by uniaxial stress reduces the hybridization gap, triggering thereby magnetic order and an insulator-to-metal transition (Supplemental Material~\cite{supp}); this is an interesting proposal for experiment. In fact, the $\Gamma_8$ quartet splitting is naturally present near the surface as well as near defects, and should be relevant for the analysis of the surface states and transport properties of SmB$_6$. For instance, the surface magnetism reported in SmB$_6$~\cite{Nak16,Ais22} can be understood as a natural consequence of  the cubic symmetry breaking near the surface.

\begin{figure}
\centering
\includegraphics[width = 0.99 \linewidth]{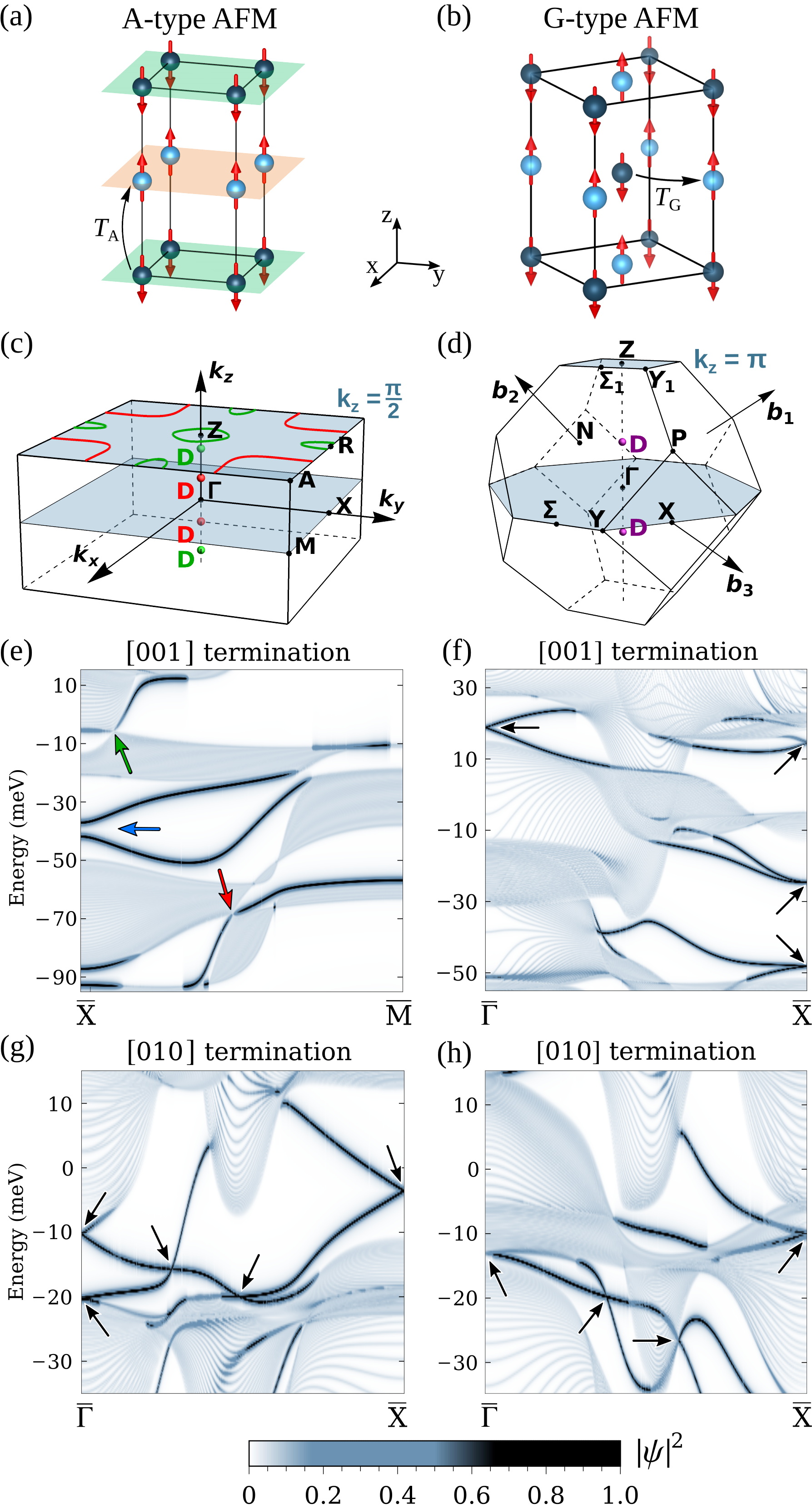}
\caption{Band topology of SmB$_6$ at valence $v=2.74$ for the A-type (left) and G-type (right) antiferromagnetic states.
(a),(b) Magnetic structures with moments indicated by the red arrows.
(c),(d) Nodal lines and Dirac points in the vicinity (green) and below (red \& purple) the Fermi level.
(e-h) Band structures of a 100-layer slab with [001] and [010] termination, respectively. The color scale represents the wave function amplitude of the 10 topmost layers. The arrows mark characteristic features of the different band topologies. }
\label{fig:3}
\end{figure}

\emph{Topological properties.}---Paramagnetic SmB$_6$ is a topological insulator with strong and weak $\mathbb{Z}_2$ invariants and mirror Chern numbers~\cite{Dze10}. Here, we present a topological analysis of the magnetically ordered metallic phases, based upon the $\Gamma_8$ quartet model of SmB$_6$. Specifically, we consider the two possible magnetic states, namely the A-type and G-type AFM orders shown in Figs.~\ref{fig:3}(a,b). Our topological analysis is based on symmetries, and therefore applies also, at least qualitatively, to similar  antiferromagnetic systems.

In the magnetic phases, the symmorphic symmetries $Pm\bar{3}m$ of the paramagnetic phase are lowered to \textit{nonsymmorphic} magnetic space groups (MSG)~\cite{Gal12}. For the A-type order we find the tetragonal MSG P$_{2c}$4/mm'm', while for the G-type order the MSG is  P$_I$4/mm'm', which is body-centered. Here we adopt the OG convention \cite{Gal12}, for which the coordinate basis of the MSG coincides with the parent SG. Both MSGs contain magnetic translations, namely $T_\mathrm{A} =T \, t(0,0,1)$ along the $z$ axis for the A-type order and $T_\mathrm{G}  = T \, t(1,0,0)$ along the $x$ axis for the G-type order, where $T = i \sigma_y K$ is the time-reversal operator. We note that $T_\mathrm{G}$ is equivalent to  $T_\mathrm{A}$ up to a lattice vector of the G-type structure. Yet, the converse is not true, i.e., $T_\mathrm{A}$ is not equivalent to $T_\mathrm{G}$ up to a lattice vector of the A-type lattice.

To study the band topologies of the AFM phases, we compute the surface states and the associated topological numbers in our model, whereby we implement the coupling of the charge carriers to the AFM texture by an on-site Zeeman term. We find that for both AFM orders the bands are Kramers degenerate, due to the presence of both magnetic translations ($T_\mathrm{A}$, $T_\mathrm{G}$) and inversion ($P$) with ${(PT_\mathrm{A})^2 = (PT_\mathrm{G})^2 = - 1}$. Moreover, both A- and G-type orders exhibit Dirac points below the Fermi level, while the A-type order has in addition nodal lines~\cite{supp}. We now discuss these two AFM orders in more detail.

(i) The A-type order exhibits Dirac nodal lines in the $k_z=\tfrac{\pi}{2}$ plane for a large parameter range (due to the wide $d$-band), see Fig.~\ref{fig:3}(c). These nodal lines are classified as type~II~\cite{Li18}, since they are strongly tilted. The appearance of Dirac nodal lines upon entering the A-type order defies the expectation that with less symmetries there are less symmetry-protected crossings. These Dirac nodal lines are protected by the $m_z$ mirror symmetry together with  $PT_A$. In the $k_z= \tfrac{\pi}{2}$ mirror plane  $PT_\mathrm{A}$ pairs identical representations of $m_z$, such that two Kramers pairs with distinct $m_z$ representations can cross in protected nodal lines. These nodal lines can be thought of as   generalizations of lines protected by inversion and off-centered mirror symmetry~\cite{Yan17,zhang_PRMat18}.

By the bulk-boundary correspondence, one expects that these nodal lines lead to states localized on the [001] surface. Indeed, Fig.~\ref{fig:3}(e) shows nodal line surface states for the red and green nodal lines of Fig.~\ref{fig:3}(c), which however partially merge with bulk bands. In order to characterize the topology of these nodal lines and the associated surface states, one might consider a mirror Berry phase \cite{Zha13,chan_drumhead_16}. However, this quantity is unsuitable, because the two orthogonal eigenspaces of the glide mirror symmetries $\tilde{m}_x$ and $\tilde{m}_y$ exchange along the $k_z$ direction. Instead, we develop a ``glide-mirror graded" Wilson loop~\cite{supp}. The spectrum of this quantity, which resolves the band degeneracy caused by $PT_\mathrm{A}$, exhibits a qualitative change at the nodal lines, i.e., the emergence of flat bands. Finally, on the [001] surface there appears also a gapped Dirac cone (blue arrow), which can be viewed as a remainder of the paramagnetic surface states, gapped by the magnetic order. While this is similar to an axion insulator \cite{Mon10, liu20}, a half-integer Hall effect is not expected because of the gapless band structure.

On the [010] surface, on the other hand, there appear no nodal line surface states but only Dirac cones, indicated by black arrows in Fig.~\ref{fig:3}(g). These Dirac cones, found also in a band structure study~\cite{Cha18} of the A-type order, are protected by mirror-Chern numbers and weak $\mathbb{Z}_2$ invariants defined within the $k_z=0$ plane~\cite{supp}.

(ii) The G-type order does not have any nodal lines, but only Dirac cones on the $\Gamma$-Z axis protected by the fourfold rotation, see Fig.~\ref{fig:3}(d). Similar to the A-type order, the G-type order has weak $\mathbb{Z}_2$ invariants in two-dimensional subsystems where $T_\mathrm{G}^2 = -1$ \cite{Chi16}. In particular, the three planes $k_x=0$, $k_y=0$, and $k_z=0$, which each contain the time-reversal invariant momenta $\Gamma$, X, and Z, have nontrivial weak invariants~\cite{supp}. By the bulk-boundary correspondence, these weak invariants lead to Dirac cone surface states for the [100], [010], and [001] terminations, see black arrows in Figs.~\ref{fig:3}(f) and~\ref{fig:3}(h).

Recent analysis of the quantum oscillation experiments at high magnetic fields suggested the presence of asymmetric nodal lines~\cite{Har18}. Motivated by this proposal, we consider the possibility of nodal lines in the high-field phase of SmB$_6$, where the moments are aligned ferromagnetically. In this state, any mirror plane may contain protected accidental nodal lines. Indeed, for the [001] alignment of the magnetic moments we find such nodal lines within the $k_z = 0$ and $k_z = \pi$ mirror planes, in close proximity to the Fermi level. These nodal lines are protected by a quantized $\pi$-Berry phase~\cite{supp}.

In conclusion, we have developed a theory for the insulator-to-magnetic-metal transition in SmB$_6$. We describe this transition as the softening of an in-gap spin exciton mode, which condenses into an AFM order at the critical density of magnetic Sm$^{3+}$ ions. The AFM order induces a simultaneous insulator-to-metal transition as well as a change in the band topology. Within the $\Gamma_8$ quartet model, relevant to SmB$_6$, we find that the band structure is transformed from a cubic topological insulator to a tetragonal magnetic metal with symmetry protected Dirac points (G-type order) or Dirac points and nodal lines (A-type order). From our theory, we expect that the magnetic metal exhibits interesting transport characteristics, for example, light- or strain-induced anomalous Hall currents in the A-type order~\cite{Rui18,fujimoto_schnyder_PRR_22,oka_light_hall_graphene}. We hope that our findings will spur experimentalists to look for these intriguing properties, and to further characterize the insulator-to-magnetic-metal transition which can be induced in SmB$_6$ by pressure or uniaxial strain.

\vspace{1em}

We thank Andreas Leonhardt for discussions. G.Kh. thanks the Max Planck-UBC-UTokyo Centre for Quantum Materials at the University of British Columbia, where this work was partially done, for hospitality. H.L. acknowledges support by the W$\ddot{\rm u}$rzburg-Dresden Cluster of Excellence on Complexity and Topology in Quantum Matter \emph{ct.qmat} (EXC 2147, project ID  390858490). A.P.S. thanks the YITP Kyoto for hospitality.

\vspace{1em}

H.L. and M.M.H. contributed equally to this work.


\clearpage
\onecolumngrid

\begin{center}
\textbf{\large Supplemental Material for\\
Correlation Induced Magnetic Topological Phases in Mixed-Valence Compound SmB$_6$}
\end{center}

\setcounter{equation}{0}
\setcounter{figure}{0}
\setcounter{table}{0}
\setcounter{page}{7}
\makeatletter
\renewcommand{\thesection}{S\Roman{section}}
\renewcommand{\thetable}{S\arabic{table}}
\renewcommand{\theequation}{S\arabic{equation}}
\renewcommand{\thefigure}{S\arabic{figure}}

Here we present the details of the electronic model (Sec.~\ref{SecTBModel}), calculation of the magnetic susceptibility (Sec.~\ref{sec_S2_mag_suscept}), and consider tetragonal crystal field effect on the phase transition (Sec.~\ref{sec_S3_TetraCF}). In Sec.~\ref{SecTopMethod}, we discuss suitable topological invariants and introduce a glide-mirror graded Wilson loop. Using these invariants, we analyze in Sec.~\ref{SecTopAnalysis} the topological properties of SmB$_6$ in different  phases.

\section{Valence states and the tight-binding model}
\label{SecTBModel}
\begin{figure}[h]
\begin{center}
\includegraphics[width=10.5cm]{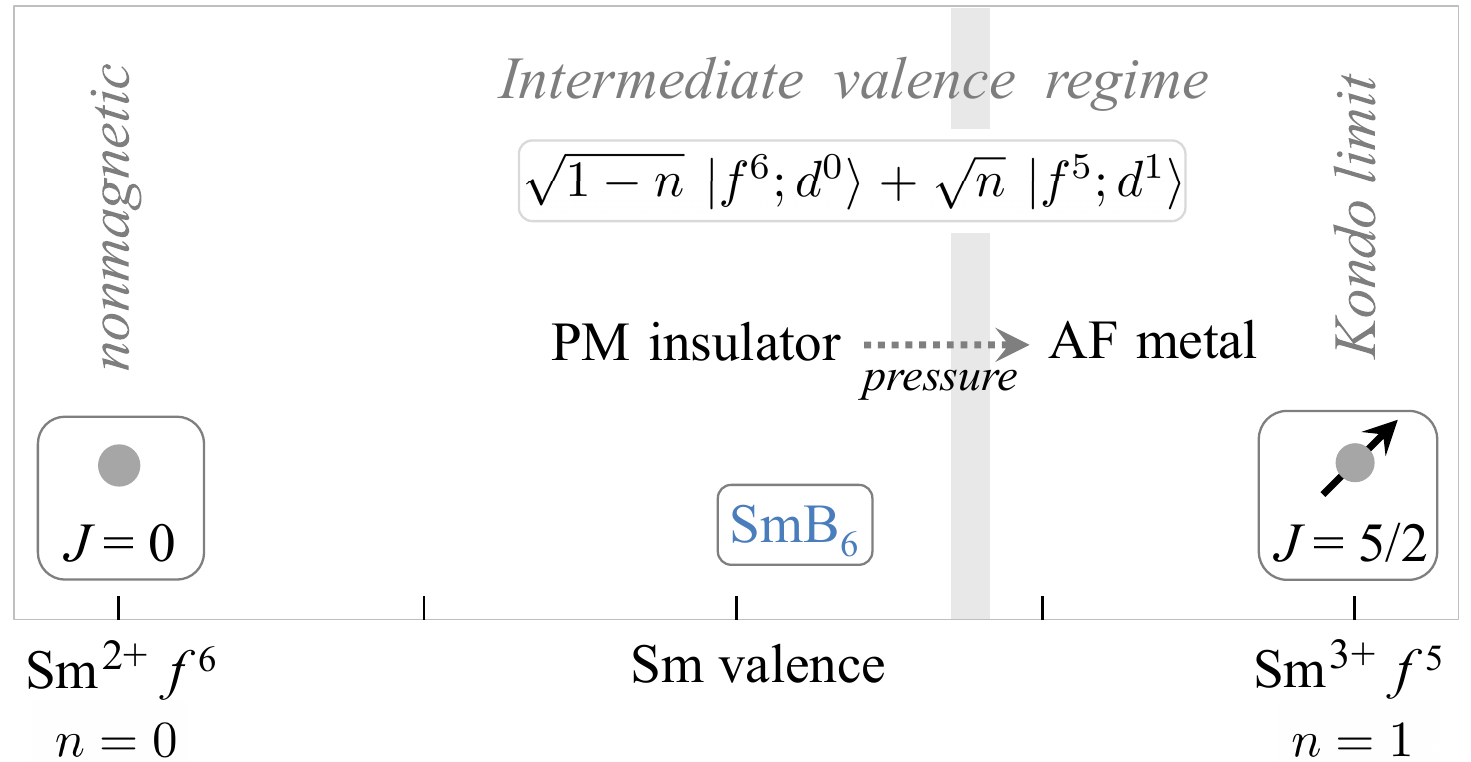}
\caption{Schematic of the lowest spin-orbit $J$ multiplets of Sm$^{2+}$ and Sm$^{3+}$ ions in SmB$_6$. These two valence states are almost evenly presented in the ground state wavefunction, with the average densities $1-n \simeq 0.44$ and $n \simeq 0.56$, respectively. As the density of magnetic ions Sm$^{3+}$ is slightly increased (to $n \sim 0.7$~\cite{But16}) under pressure, a first order phase transition from nonmagnetic-insulator to magnetic-metal takes place.
}
\label{fig:s1}
\end{center}
\end{figure}

SmB$_6$ is a prototypical intermediate valence compound, where the ionic charge configuration fluctuates between two different valence states, which are energetically close and coherently superposed in the ground state wavefunction. As illustrated in Fig.~\ref{fig:s1}, the average Sm valence in SmB$_6$ is ``in-between'' the divalent and trivalent states, and it can be tuned by external or chemical pressure thanks to the different radii of Sm$^{2+}(4f^6)$ and Sm$^{3+}(4f^5)$ ions, the latter being smaller because of the smaller number of $4f$ electrons.

Microscopically, the valence fluctuations are caused by hybridization of the $4f$ electrons with $5d$ conduction band states, a process which can schematically be expressed as $f^6d^0 \Leftrightarrow f^5d^1$. Both valence states of the Sm ion, $f^6(L=3,S=3)$ and $f^5(L=5,S=5/2)$, have a number of spin-orbit multiplet levels with total angular momentum $J=0, 1, ...$ and $J=5/2, 7/2, ...$, respectively. The general Hamiltonian including all these states is obviously complex, but the transitions between the lowest $f^6(J=0)$ singlet and $f^5(J=5/2)$ states of interest can be described as the removal/addition of an $f$-electron with the angular momentum $j=5/2$. In other words, thanks to the trivial $J=0$ ground state of Sm$^{2+}$ ion, we can express the valence fluctuations in terms of the hoppings of a single $f$ electron, with the effective amplitudes reduced by the coefficients of fractional parentage~\cite{Saw16}. As we neglect the higher-lying $\Gamma_7$ level of Sm$^{3+}$~\cite{Amo19} in our model, the $f(j=5/2)$ electron has to be in the $\Gamma_8$ state as well.

The $\Gamma_8$ quartet of $f$ electron can be represented by two Kramers doublets, which we denote by $\alpha$ and $\beta$; their wave functions, written in terms of the $j=5/2$ angular momentum projections, read as follows:
\begin{align}
\label{eq:GS8}
|\alpha_{\pm}\rangle = \left |\pm\tfrac{1}{2}\right \rangle, \ \ \ \ \ \ \ \
|\beta_{\pm}\rangle = \sqrt{\tfrac{5}{6}}\left |\pm\tfrac{5}{2}\right \rangle
+ \sqrt{\tfrac{1}{6}} \left  |\mp\tfrac{3}{2} \right \rangle,
\end{align}
where the subscript $\pm$ denotes the pseudospin-1/2 index. To calculate hopping integrals between these states and $d$ orbitals, one actually needs to express the above $\left |j_z\right \rangle$ functions in terms of the $f$ orbital $l=3$ and spin $s=1/2$ angular momentum projections $|l_z,s_z\rangle$:
\begin{align}
\left |\tfrac{5}{2}\right \rangle &
= \sqrt{\tfrac{6}{7}} \; \left |3, -\tfrac{1}{2}\right \rangle
 -\sqrt{\tfrac{1}{7}}  \; \left |2, \tfrac{1}{2} \right \rangle  \;,
\  \  \  \
\left |-\tfrac{5}{2}\right \rangle =
-\sqrt{\tfrac{6}{7}} \; \left |-3, \tfrac{1}{2} \right \rangle
+\sqrt{\tfrac{1}{7}}  \; \left |-2, -\tfrac{1}{2} \right \rangle  \;,
\notag\\
\left |\tfrac{3}{2}\right \rangle &
= \sqrt{\tfrac{5}{7}} \; \left |2, -\tfrac{1}{2}\right \rangle
 -\sqrt{\tfrac{2}{7}} \; \left |1, \tfrac{1}{2}\right \rangle \;,
\  \  \  \
\left |-\tfrac{3}{2}\right \rangle =
-\sqrt{\tfrac{5}{7}} \; \left |-2, \tfrac{1}{2}\right \rangle
 +\sqrt{\tfrac{2}{7}}  \; \left |-1, -\tfrac{1}{2}\right \rangle \;,
\notag\\
\left |\tfrac{1}{2}\right \rangle &
= \sqrt{\tfrac{4}{7}} \; \left |1, -\tfrac{1}{2}\right \rangle
 -\sqrt{\tfrac{3}{7}}  \; \left |0, \tfrac{1}{2}\right \rangle \;,
\  \  \  \
\left |-\tfrac{1}{2}\right \rangle =
-\sqrt{\tfrac{4}{7}} \; \left |-1, \tfrac{1}{2}\right \rangle
 +\sqrt{\tfrac{3}{7}}  \; \left |0, -\tfrac{1}{2}\right \rangle \;.
 \label {eq:wf_f}
\end{align}

Using the above $\Gamma_8$ quartet and two $5d$ orbitals of $e_g$ symmetry ($d_{z^2}$ and $d_{x^2-y^2}$), we first derive the spin-orbital structure of the hopping integrals between these states up to second nearest-neighbour bonds. This is done following the Slater-Koster tables for the $f$ and $d$ orbitals~\cite{Sla54,Tak80}. Then we transform the Hamiltonian into the momentum space, and represent it in the following matrix form, written in the basis of the $\{\alpha_+,\alpha_-,\beta_+,\beta_-,d_{z^2 \uparrow},d_{z^2 \downarrow},d_{x^2-y^2 \uparrow},d_{x^2-y^2 \downarrow}\}$ states:
\begin{align}
\mathcal{H}=\left[ \begin{array}{cc}
E_f I +  H_{ff} &  H_{df} \\
  H_{df}^{\ast} & E_d I + H_{dd}\end{array}\right] .
\label{eq:H}
\end{align}
Here, $E_f$ and $E_d$ are the on-site energies for $f$ and $d$ orbitals, and $I$ is the $4 \times 4$ unit matrix. The $4 \times 4$ block matrices $H_{ff}= H_{ff1}+H_{ff2}$ and $H_{dd}= H_{dd1}+H_{dd2}$ are the kinetic energies for $f$ and $d$ electrons, respectively, with the subscripts $1$ and $2$ representing the nearest-neighbor (NN) and the next-nearest-neighbor (NNN) bonds; the $4 \times 4$ matrix $H_{df}=H_{df1}+H_{df2}$ denotes the hybridization between $d$ and $f$ orbitals. The explicit forms of these submatrices are introduced term by term below.

The NN hopping energy $H_{ff1}$ for the $\Gamma_8$ states $\{\alpha_+,\alpha_-,\beta_+,\beta_-\}$ reads as:
\begin{align}
H_{ff1}=
\left[ \begin{array}{cccc}
t_{f1}\phi_1(\vc k) +t'_{f1}\phi_2(\vc k)
& 0 & (t_{f1}-t'_{f1})\phi_3(\vc k)  & 0  \\
0 & t_{f1}\phi_1(\vc k) +t'_{f1}\phi_2(\vc k) & 0  &  (t_{f1}-t'_{f1})\phi_3(\vc k)  \\
(t_{f1}-t'_{f1})\phi_3(\vc k) & 0 &  t_{f1}\phi_2(\vc k) +t'_{f1}\phi_1(\vc k) & 0 \\
0 & (t_{f1}-t'_{f1})\phi_3(\vc k) &  0  & t_{f1}\phi_2(\vc k) +t'_{f1}\phi_1(\vc k)
\end{array}\right] ,
\end{align}
where $t_{f1}$ ($t'_{f1}$) is an effective, i.e., rescaled by the coefficients of fractional parentage~\cite{Saw16}, hopping between $\alpha$-$\alpha$ ($\beta$-$\beta$) orbitals along the $z$-axis. The real functions $\phi_i$ are given by: $\phi_1(\vc k)=c_x+c_y+4c_z$, $\phi_2(\vc k)=3(c_x+c_y)$, and $\phi_3(\vc k)=\sqrt{3}(c_y-c_x)$, with $c_j = \cos k_j~(j=x,y,z)$.

For the NNN hopping terms between the $f$ orbitals, we have
\begin{align}
H_{ff2}\!=\!
{ \small
\left[ \begin{array}{cccc}
t_{f2}\phi_4(\vc k) +t'_{f2}\phi_5(\vc k)
& 0 & (t_{f2}\!-\!t'_{f2})\phi_6(\vc k)\!+\! it''_{f2}\phi_7(\vc k) & t''_{f2}\phi_8(\vc k)  \\
0 & t_{f2}\phi_4(\vc k) +t'_{f2}\phi_5(\vc k) & -t''_{f2}\phi^{\ast}_8(\vc k)  &  (t_{f2}\!-\!t'_{f2})\phi_6(\vc k)\!-\! it''_{f2}\phi_7(\vc k)  \\
(t_{f2}\!-\!t'_{f2})\phi_6(\vc k)\!-\! it''_{f2}\phi_7(\vc k)  & -t''_{f2}\phi_8(\vc k) &  t_{f2}\phi_5(\vc k) +t'_{f2}\phi_4(\vc k) & 0 \\
t''_{f2}\phi^{\ast}_8(\vc k) & (t_{f2}\!-\!t'_{f2})\phi_6(\vc k)\!+\! it''_{f2}\phi_7(\vc k) &  0  & t_{f2}\phi_5(\vc k) +t'_{f2}\phi_4(\vc k)
\end{array}\right].
}
\end{align}
Here, $\phi_4(\vc k)=c_xc_z+c_yc_z+4c_xc_y$, $\phi_5(\vc k) =3c_z(c_x+c_y)$,
$\phi_6(\vc k) =\sqrt{3}(c_x-c_y)c_z$, $\phi_7(\vc k) =4s_x s_y$, $\phi_8(\vc k) =4(s_x+is_y)s_z$, and $s_j=\sin k_j~(j=x,y,z)$. $t_{f2}$ ($t'_{f2}$, $t''_{f2}$) is the effective hopping amplitude between in-plane NNN $\alpha$-$\alpha$ ($\beta$-$\beta$, $\alpha$-$\beta$) orbitals.

Correspondingly, the matrix forms for NN hopping $H_{dd1}$ and NNN hopping $H_{dd2}$ between $d$ orbitals, written in the $\{ d_{z^2 \uparrow},d_{z^2 \downarrow},d_{x^2-y^2 \uparrow},d_{x^2-y^2 \downarrow} \}$ basis, with $ \uparrow,\downarrow$ representing the up and down spin states, are:
\begin{align}
H_{dd1}=
\left[ \begin{array}{cccc}
t_{d1}\phi_1(\vc k) +t'_{d1}\phi_2(\vc k)
& 0 & (t_{d1}-t'_{d1})\phi_3(\vc k)  & 0  \\
0 & t_{d1}\phi_1(\vc k) +t'_{d1}\phi_2(\vc k) & 0  &  (t_{d1}-t'_{d1})\phi_3(\vc k)  \\
(t_{d1}-t'_{d1})\phi_3(\vc k) & 0 &  t_{d1}\phi_2(\vc k) +t'_{d1}\phi_1(\vc k) & 0 \\
0 & (t_{d1}-t'_{d1})\phi_3(\vc k) &  0  & t_{d1}\phi_2(\vc k) +t'_{d1}\phi_1(\vc k)
\end{array}\right] ,
\end{align}
and
\begin{align}
H_{dd2}=
{ \small
\left[ \begin{array}{cccc}
t_{d2}\phi_4(\vc k) +t'_{d2}\phi_5(\vc k)
& 0 & (t_{d2}-t'_{d2})\phi_6(\vc k)  & 0 \\
0 & t_{d2}\phi_4(\vc k) +t'_{d2}\phi_5(\vc k) & 0  &  (t_{d2}-t'_{d2})\phi_6(\vc k)   \\
(t_{d2}-t'_{d2})\phi_6(\vc k)   & 0 &  t_{d2}\phi_5(\vc k) +t'_{d2}\phi_4(\vc k) & 0 \\
0& (t_{d2}-t'_{d2})\phi_6(\vc k) &  0  & t_{d2}\phi_5(\vc k) +t'_{d2}\phi_4(\vc k)
\end{array}\right] .
}
\end{align}

The NN hybridization term $H_{df1}$ between $d$- and $f$-states reads as:
\begin{align}
H_{df1}=i
\left[ \begin{array}{cccc}
V_1 \phi_9 (\vc k)  & -(V_1+3V'_1) \phi_{10}(\vc k) &0& \sqrt{3}(V_1-V'_1)\phi^{\ast}_{10}(\vc k) \\
-(V_1+3V'_1) \phi^{\ast}_{10}(\vc k)& - V_1\phi_9(\vc k) &\sqrt{3}(V_1-V'_1) \phi_{10}(\vc k)  &  0\\
0&\sqrt{3}(V_1-V'_1) \phi^{\ast}_{10}(\vc k) &V'_1 \phi_9(\vc k)  & -(3V_1+V'_1) \phi_{10}(\vc k) \\
\sqrt{3} (V_1-V'_1) \phi_{10}(\vc k) & 0& -(3V_1+V'_1) \phi^{\ast}_{10}(\vc k)   & -V'_1\phi_9(\vc k)
\end{array}\right],
\end{align}
where $V_1$ ($V'_1$) represents an effective hybridization amplitude between $\alpha$-$d_{z^2}$ ($\beta$-$d_{x^2-y^2}$) orbitals, $\phi_9(\vc k) =-4s_z$ and $\phi_{10}(\vc k) =s_x-is_y$.

Finally, the NNN hybridization Hamiltonian $H_{df2}$ is:
\begin{align}
H_{df2}=i
{
\small
\left[ \begin{array}{cc}
[V_2+3V'_2-\sqrt{3}(V''_2+V'''_2)] \phi_{11}(\vc k)
& V_2 \phi_{13}(\vc k) +[V_2+3V'_2+\sqrt{3}(V''_2+V'''_2)] \phi_{14}(\vc k)  \\
V_2 \phi^{\ast}_{13}(\vc k)+[V_2+3V'_2+\sqrt{3}(V''_2+V'''_2)] \phi^{\ast}_{14}(\vc k)
&  -[V_2+3V'_2-\sqrt{3}(V''_2+V'''_2)] \phi_{11}(\vc k)\\
\ [\sqrt{3}(V_2-V'_2)-3V''_2+V'''_2]\phi_{12}(\vc k)
& V'''_2 \phi^{\ast}_{13}(\vc k) +[\sqrt{3}(V_2-V'_2)+3V''_2-V'''_2] \phi^{\ast}_{14}(\vc k)     \\
V'''_2 \phi_{13}(\vc k) +[\sqrt{3}(V_2-V'_2)+3V''_2-V'''_2] \phi_{14}(\vc k)
& -[\sqrt{3}(V_2-V'_2)-3V''_2+V'''_2]\phi_{12}(\vc k)
\end{array}\right.
}
\notag \\
\notag \\
\left. \begin{array}{cc}
[\sqrt{3}(V_2-V'_2)+V''_2-3V'''_2] \phi_{12}(\vc k)
& V''_2 \phi^{\ast}_{13}(\vc k) +[\sqrt{3}(V_2-V'_2)-V''_2+3V'''_2] \phi_{14}^{\ast}(\vc k)  \\
 V''_2 \phi_{13}(\vc k) +[\sqrt{3}(V_2-V'_2)-V''_2+3V'''_2] \phi_{14}(\vc k)
&  -[\sqrt{3}(V_2-V'_2)+V''_2-3V'''_2] \phi_{12}(\vc k)\\
\ [3V_2+V'_2+\sqrt{3}(V''_2+V'''_2)]\phi_{11}(\vc k)
& V'_2 \phi_{13}(\vc k) +[3V_2+V'_2-\sqrt{3}(V''_2+V'''_2)] \phi_{14}(\vc k)     \\
V'_2 \phi^{\ast}_{13}(\vc k) +[3V_2+V'_2-\sqrt{3}(V''_2+V'''_2)] \phi^{\ast}_{14}(\vc k)
& -[3V_2+V'_2+\sqrt{3}(V''_2+V'''_2)]\phi_{11}(\vc k)
\end{array}\right] ,
\end{align}
where $V_2$ ($V'_2$, $V''_2$, $V'''_2$) is the hybridization between in-plane NNN $\alpha$-$d_{z^2}$ ($\beta$-$d_{x^2-y^2}$, $\alpha$-$d_{x^2-y^2}$, $\beta$-$d_{z^2}$) orbitals, and $\phi_{11}(\vc k)=s_z(c_x+c_y)$,  $\phi_{12}(\vc k)=s_z(c_x-c_y)$, $\phi_{13}(\vc k)=4(s_xc_y-ic_xs_y)$, and $\phi_{14}(\vc k)=c_z(s_x-is_y)$.
We note that the spin-orbital structure of the hopping Hamiltonians obtained above is consistent with the previous works~\cite{Tak11,Bar14}.

\begin{figure}
\begin{center}
\includegraphics[width=10.5cm]{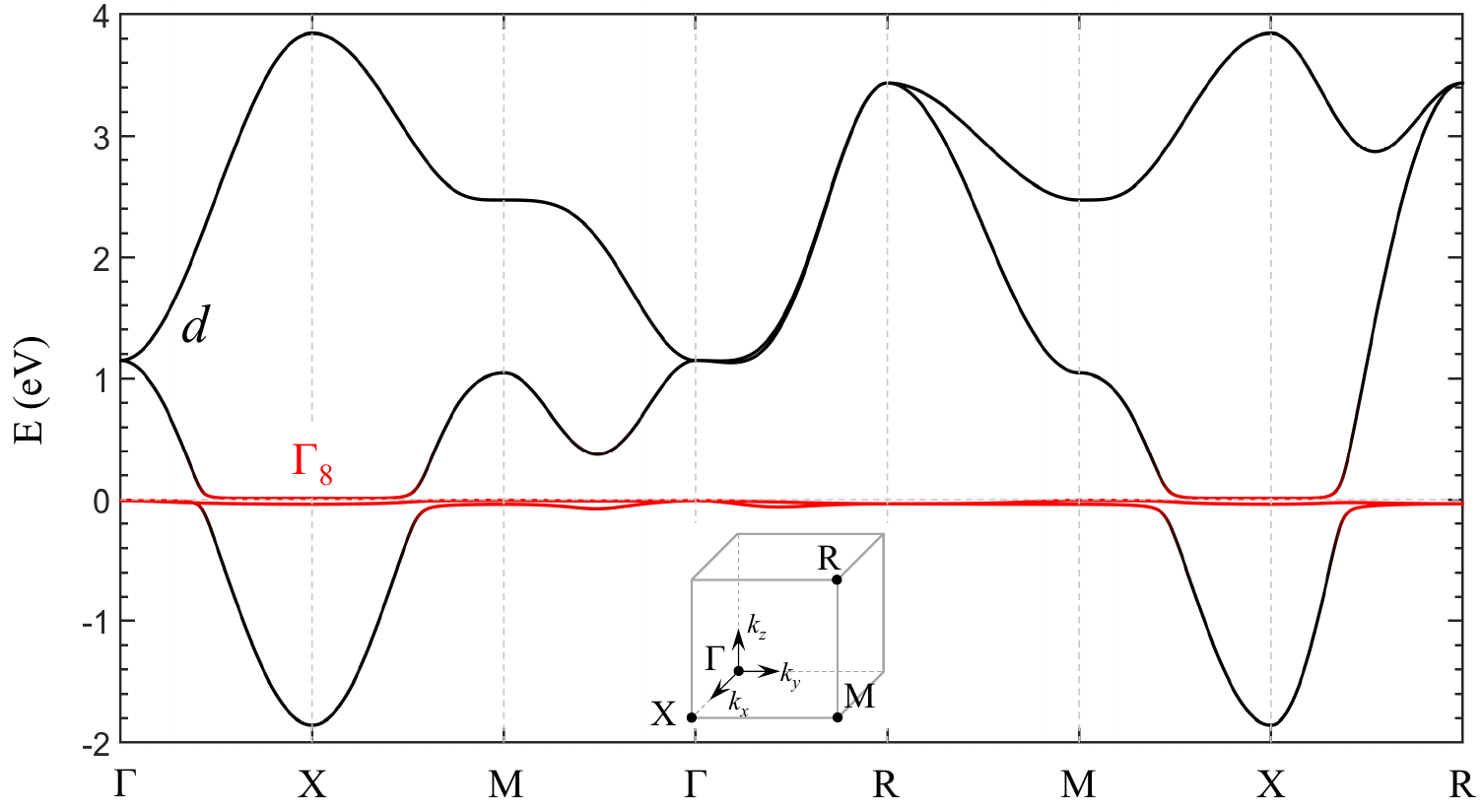}
\caption{Band structure along high-symmetry paths in the Brillouin zone, calculated with the parameters given in Eq.~\eqref{eq:para}. Black and red colors are for $d$- and $f$-like band states.}
\label{fig:s2}
\end{center}
\end{figure}

For the valence state with $v=2.56$, i.e., at ambient pressure, we use the following parameter set:
\begin{alignat}{7}
E_d&=1.6 ~{\rm eV} \;,       &\ \ \  E_f&=-0.02 ~{\rm eV}\;,  & \ \ \ \ \ \ & \ \ \ \ \ \ & \ \ \ \ \ \ \ & \ \ \ \ \ \ & \ \ \ \ \ \ \ & \ \ \ \ \ \ & \ \ \ \ \ \ \ \ & \ \ \ \ \ \
\notag \\
t_{d1}&=-0.32 ~{\rm eV} \;,  &  t'_{d1}&=-0.4t_{d1} \;, &  t_{d2}&=-0.6 t_{d1} \;,  & t'_{d2}&=-0.4t_{d2} \;, &  &   & &
\notag \\
t_{f1}&=0.0036 ~{\rm eV} \;, & t'_{f1}&=-0.4t_{f1}  \;,  & t_{f2}&=-0.2 t_{f1}  \;,  &  t'_{f2}&=-0.4t_{f2}  \;,  & t''_{f2}&=0.4t_{f2} \;, & &
\notag \\
V_1&= 0.036 ~{\rm eV}  \;,   &  V'_1&=-0.4 V_1 \;,       &   V_2&=0.1 V_1 \;,  &  V'_2&=-0.2V_2  \;,  &  V''_2&=-0.2V_2 \;, &   V'''_2&=0.1V_2 \; .
\label{eq:para}
\end{alignat}
The resulting band structure is shown in Fig.~\ref{fig:s2}. The $4f$-dominated bands are located around the Fermi level, while the $5d$-like states disperse widely from $-2$~eV to $4$~eV. A hybridization gap of $\simeq 17$~meV is consistent with the ARPES data~\cite{Neu13}.

\section{Magnetic susceptibility and spin exciton mode}
\label{sec_S2_mag_suscept}
The bare magnetic susceptibility $\chi_0({\bm q},\omega)$ in Eq.~(3) of the main text is evaluated as follows:
\begin{align}
\chi_0({\bm q}, \omega ) \!=\! \sum_{{\bm k}lm} \;
\frac{|\langle l_{{\bm k}} |J_z |m_{{\bm k}+{\bm q}} \rangle|^2}
  {\omega \!+\! E_{l,\vc k} \!-\! E_{m, \vc{k}+\vc q}\!+\!i0^+} \;
  (f_{E_{m, \vc{k}+\vc q}}\!-\!f_{E_{l,\vc k}}).
\label{eq:chi0}
\end{align}
Here, $l$ and $m$ run over the eight bands derived from the $\Gamma_8$ and $d(e_g)$ states, as discussed in the previous section. Momentum ${\bm k}$ runs over the first Brillouin zone. The $z$ component of the angular momentum $J=5/2$, projected onto the $\Gamma_8$ quartet, is given by $J_z =s_{\alpha}^z +\frac{11}{3} s_{\beta}^z$, where the pseudospins one-half $s_{\alpha}$ and $s_{\beta}$ operate within the $\alpha$ and $\beta$ Kramers doublets defined in Eq.~\eqref{eq:GS8} above.

\begin{figure}
\begin{center}
\includegraphics[width=13.5cm]{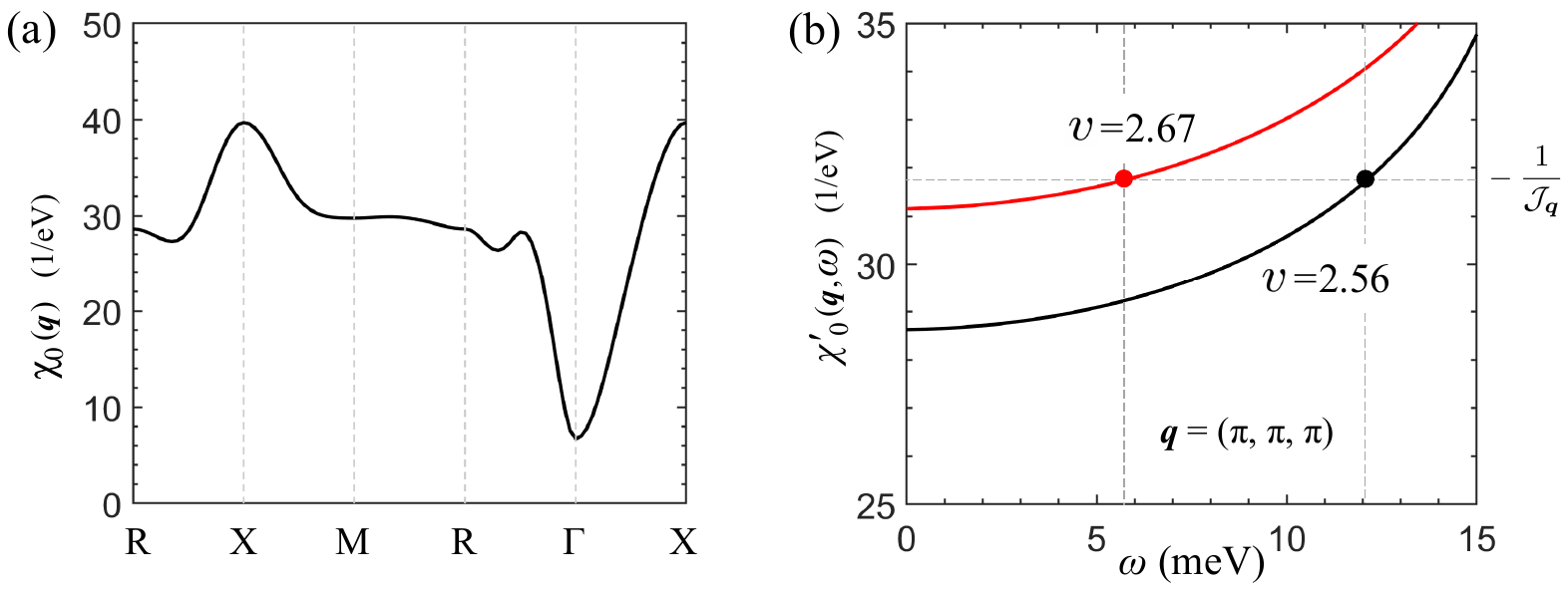}
\caption{(a) Static magnetic susceptibility $\chi_0(\vc q,0)$ at valence $v=2.56$. (b) Frequency dependence of the real part of the bare magnetic susceptibility $\chi'_0(\vc q,\omega)$ for R-point with $\vc q=(\pi,\pi,\pi)$, at valence $v=2.56$ (black) and $v=2.67$ (red). Black and red dots determine the corresponding  spin-exciton energies $\omega_{\bf q}$ following from Eq.~\eqref{eq:wq}. The exchange couplings $(\mathcal{J}_1, \mathcal{J}_2, \mathcal{J}_3) = (14.7, 6.5, 2.7)$~meV are used as in the main text. Note that $\omega_{\bf q}$ is reduced by a factor of two as the Sm valence $v$ is slightly increased.
}
\label{fig:s3}
\end{center}
\end{figure}

Figure~\ref{fig:s3}(a) shows the static susceptibility $\chi_0(\vc q)$ for $v=2.56$, calculated using the parameter set of Eq.~\eqref{eq:para}. Its momentum dependence is decided by the heavy-band dispersions of $\Gamma_8$ character, and has a maximum at the X-point with $\vc q = (\pi,0,0)$. However, the exchange couplings between the $\Gamma_8$ states must be considered to obtain the full momentum and energy dependence of the magnetic response. Within the random phase approximation, one finds
\begin{align}
\chi( {\bm q},\omega)=\frac{\chi_0( {\bm q},\omega)}{1+\mathcal{J}_{\bm q} \chi_0({\bm q},\omega)} \;.
\label{eq:chi}
\end{align}
Including the exchange couplings $\mathcal{J}_R$ up to third-nearest neighbours, we obtain
\begin{align}
\mathcal{J}_{\bm q} =2 \mathcal{J}_1 (c_x+c_y+c_z)
+4 \mathcal{J}_2 (c_xc_y+c_yc_z+c_zc_x) +8 \mathcal{J}_3 c_xc_yc_z,
 \end{align}
where $c_j = \cos q_j~(j=x,y,z)$.
%
%
In addition to particle-hole excitations from $\chi_0({\bm q},\omega)$, Eq.~\eqref{eq:chi} contains a pole at energy $\omega_{\vc q}$, which satisfies the following equation:
\begin{align}
  \chi'_0({\bm q},\omega_{\vc q}) = -\frac{1}{\mathcal{J}_{\bm q}} \;.
  \label{eq:wq}
\end{align}
The spectral weight $A_{\vc q}$ of this collective (``spin-exciton'') mode can be obtained by expanding Eq.~\eqref{eq:chi} near the pole. It is proportional to the inverse of the derivative of the real part of bare susceptibility at the peak position $\omega_{\vc q}$:
\begin{align}
A_{\vc q}=\frac{1}{\mathcal{J}_{\bm q}^2} \left( \frac{d \chi'_0( {\bm q},\omega) }{d \omega} \right)_{\omega=\omega_{\vc q}} ^{-1}.
\label{eq:int}
\end{align}
As illustrated in Fig.~\ref{fig:s3}(b), the susceptibility $\chi'_0({\bm q},\omega)$ increases with the valence $v$, resulting in a strong softening of the spin-exciton energy. Consequently, the spin-exciton peak gains a large intensity, see Fig.~2(a) of the main text.

A comment is in order concerning the spin susceptibility calculations. It is essential to use a realistic $\Gamma_8$ quartet model, because it contains an additional flat band near the Fermi level, in comparison with a simple $f$-doublet model often used in the literature. Our $\Gamma_8$ quartet model calculations, however, still underestimate the $\chi_0({\bm q},\omega)$ values, because we neglected transitions to the $\Gamma_7$ and higher levels. While the corresponding Van Vleck-type contributions would not affect the momentum dependences of $\chi_0({\bm q},\omega)$, thus not changing the qualitative picture, the exchange parameter values $\mathcal{J}_{\bm R}$, required to generate the spin-exciton peak positions consistent with the experimental data, will decrease accordingly.

In the magnetic phases, we treat the exchange interactions in Eq.~(2) of the main text in a standard molecular field approximation, and calculate the mean-field value of $\langle J_z\rangle$ in a self-consistent way. The ordered magnetic moment $M$, shown in Fig.~2(b) of the main text, is defined as $M = g_J \langle J_z\rangle$, with the $g$-factor of $g_J=2/7$.

\section{Tetragonal crystal field effect on the phase transition}
\label{sec_S3_TetraCF}
%
SmB$_6$ becomes magnetically ordered metal when the exchange field is strong enough to overcome the hybridization gap. This is realized under pressure, which increases the Sm valence hence the density of magnetoactive Sm$^{3+}$ ions. The present $\Gamma_8$ quartet model suggests that there is an alternative way to stabilize magnetic order even at ambient pressure, by reducing the cubic symmetry to tetragonal one, that is, by splitting $\Gamma_8$ quartet into the $\alpha$ and $\beta$ doublet states. In practice, such a tetragonal crystal field is induced by an external strain. It is important to notice, however, that cubic symmetry is actually broken at the surface of crystal even without any strain, and also near the defects in the bulk, so the tetragonal field effect considered here is generic and essential to understand the precise structure of the surface and defect states in mixed-valence compounds of cubic symmetry such as SmB$_6$.

\begin{figure}
\begin{center}
\includegraphics[width=15cm]{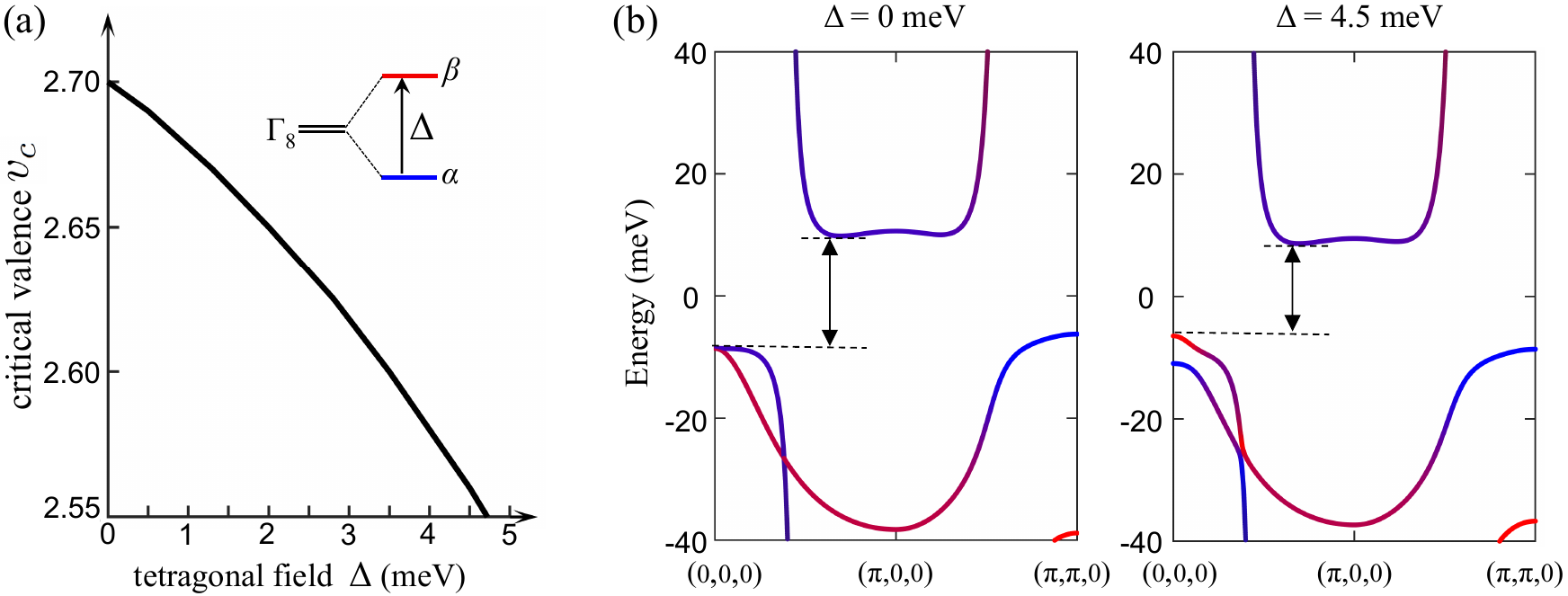}
\caption{(a) Critical valence $v_c$ for magnetic ordering in SmB$_6$ as a function of tetragonal crystal field, which splits the $\Gamma_8$ quartet into $\alpha$ and $\beta$ doublets by $\Delta$ (inset).
  (b) Band structure for the valence state with $v=2.56$ (corresponding to ambient pressure), calculated at different $\Delta$ values. The blue and red colors indicate the contributions from the $\alpha$ and $\beta$ doublets, respectively. Tetragonal splitting reduces the band gaps as indicated by double arrows.}
\label{fig:s4}
\end{center}
\end{figure}

To substantiate the above physical picture, we have calculated the critical valence $v_c$, at which the magnetic order and concomitant metal-insulator transition should appear in SmB$_6$, by including the tetragonal crystal field Hamiltonian $\Delta (n_\beta - n_\alpha)/2$. The result shown in Fig.~\ref{fig:s4}(a) tells that the tetragonal splitting as small as $\Delta\simeq4.5$ meV is sufficient to reduce the critical $v_c$ to 2.56, i.e., to the Sm valence in SmB$_6$ at ambient pressure, thus driving this mixed-valence compound into a magnetically ordered state. The origin of this effect is simple and explained in Fig.~\ref{fig:s4}(b). Namely, splitting of the $\Gamma_8$ quartet modifies the band structure near the Fermi level and reduces the hybridization gap. This enhances the magnetic response, given by Eqs.~\eqref{eq:chi0} and \eqref{eq:chi}, resulting in the spin-exciton softening and magnetic order. Physically, strong impact of the tetragonal field is due to the proximity of SmB$_6$ to a magnetic order, which can be controlled by strain and/or uniaxial pressure as suggested by our work. In fact, the tetragonal distortion which supports magnetic order is naturally present at the surface of crystal. We believe that this provides a natural explanation for the surface magnetism observed in SmB$_6$ at ambient pressure~\cite{Nak16,Ais22}.

\section{Topological invariants for materials with glide mirror symmetries}
\label{SecTopMethod}

For the paramagnetic as well as the antiferromagnetic phases of SmB$_6$ the combination of inversion symmetry $P$ with time-reversal symmetry (with the operators $T$, $T_A$, and $T_G$, respectively) leads by Kramers theorem to twofold degenerate bands.
Note that despite the translations included in $T_A$ and $T_G$ the prerequisite $(PT_A)^2 = (PT_G)^2 = -1$ of the Kramers theorem holds.
Due to the degeneracy for each electronic band, one can freely choose a basis of the corresponding two-dimensional eigenspace of the Hamiltonian.
To describe the topology of gapped subsystems, besides using the conventional $\mathbb{Z}_2$ invariant \cite{Fu07}, it is possible to resolve the internal structure of the degenerate bands by considering eigenspaces of the crystalline symmetries.
One example of this is the mirror Chern number, which refines the $\mathbb{Z}_2$ invariant \cite{Teo08}.
Hereby, one considers a mirror plane in reciprocal space, where the Hamiltonian is completely gapped.
Using the mirror symmetry one may block-diagonalize the Hamiltonian into two blocks, one for each mirror eigenvalue.
Each matrix block is found to exhibit a non-zero Chern number, whereas the total Chern number vanishes due to time-reversal symmetry (or the corresponding anti-unitary symmetry of an antiferromagnet).
A non-zero mirror Chern number is defined as the difference between the Chern numbers of the two matrix blocks.

In order to describe the topology of nodal lines, which are gapless features unlike the previously discussed gapped systems, one might try to define a mirror Berry phase in an analogous fashion  \cite{Zha13}.
However, it turns out that this is not a viable approach for the A-type antiferromagnetic phase of SmB$_6$.
The reason is that in the A-type antiferromagnetic phase the available mirror symmetries are \textit{nonsymmorphic}, e.g., $\tilde{m}_x$ is a glide mirror symmetry, since it comprises a translation $t(0,0,1)$ along the z direction.
As a result the projection into a $\tilde{m}_x$ symmetry eigenspace leads to a block-diagonal Hamiltonian that is not periodic in the Brillouin zone anymore.
Instead, at $k_z = -\pi/2$ the eigenstates of each block become pairwise orthogonal to those at $k_z = +\pi/2$.
To see this, consider the glide mirror eigenvalues $\{ \pm i \mathrm{e}^{i k_z} \}$ of $\tilde{m}_x$, which exchange over the course of the Brillouin zone.
Note that we describe the Brillouin zone of the doubled magnetic unit cell with the original lattice constants, hence the Brillouin zone extends over the interval $k_z \in [ -\pi/2, \pi/2 )$ for the A-type antiferromagnetic phase.
The exchange of mirror eigenvalues is independent of the unit cell convention and makes it impossible to consider a mirror Berry phase or, more generally, a Wilson loop in a single mirror subspace.
Nevertheless, we will show below that a \emph{mirror-graded} Wilson loop over twice the length of the Brillouin zone captures the band topology.

A non-Abelian Berry connection is needed to treat a system with degenerate bands \cite{Hat05}.
Hence, we consider the non-Abelian Wilson loop in a tight-binding formulation \cite{Ale14}
\begin{align}
	[\mathcal{W}(\mathcal{L})]^{mn}
	&=
	\left \langle U^m_{\bm{k}^{(0)} + \bm{G}}  \right |
	W(\bm{k}^{(0)} + \bm{G} \leftarrow \bm{k}^{(0)})
	\left | U^n_{\bm{k}^{(0)}}  \right \rangle,
	\label{Eq_fullTBWilsonLoop}
	\\
	W(k^{(\gamma)} \leftarrow k^{(\beta)})
	&=
	\prod_{\alpha}^{k^{(\gamma)} \leftarrow k^{(\beta)}}	
	\mathcal{P}^{\text{occ}}_{\bm{k}^{(\alpha)}},
	\\
	\mathcal{P}^{\text{occ}}_{\bm{k}}
	&=
	\sum^{n_\text{occ}}_{i = 1}
	\left | U^i_{\bm{k}} \right \rangle
	\left \langle  U^i_{\bm{k}}  \right |,
\end{align}
where $\bm{G}$ is the reciprocal lattice vector that relates two equivalent starting points $\bm{k}_0$.
The Wilson matrix $W(\mathcal{L})$, Eq.~(\ref{Eq_fullTBWilsonLoop}), is given by the product of
the projectors $\mathcal{P}^{\text{occ}}_{\bm{k}} $ on the $n_\text{occ}$ occupied eigenstates  $| U^i_{\bm{k}} \rangle $ of the Hamiltonian along a closed loop $\mathcal{L}$.
The $k$-points along the Wilson loop $\mathcal{L}$ are $\{  \bm{k}^{(0)} + \bm{G} , \bm{k}^{(N)}, \bm{k}^{(N-1)}, \ldots, \bm{k}^{(0)} \}$ with $N>1$, where a periodic gauge is assumed, i.e.,  $| U^n_{\bm{k}^{(0)} + \bm{G}}  \rangle = | U^n_{\bm{k}^{(0)}}  \rangle $.
For an arbitrary, not necessarily closed, segment of the loop $\mathcal{L}$ we denote the corresponding tight-binding Wilson matrix by $W(k^{(\gamma)} \leftarrow k^{(\beta)})$.
While the eigenspectra of the continuum Wilson loop and the tight-binding Wilson loop matrix, Eq.~(\ref{Eq_fullTBWilsonLoop}), may differ, they coincide in their real eigenvalues, and in their complex eigenvalues up to a phase \cite{Ale14}.
This turns out to be sufficient to qualitatively capture the flow of the Wilson loop spectrum as a function of $\bm{k}^0$, which in turn characterizes the topology of the surface bands \cite{Fid11}.

The bulk-boundary correspondence that relates surface bands to the bulk topology \cite{Fid11} requires the use of the cell-periodic basis convention as introduced in Ref.~\cite{Van18}.
In the paramagnetic phase, Eq.~(\ref{eq:H}), the convention plays no role, because there is only one site per unit cell, whereas in the antiferromagnetic phases there are two lattice sites per unit cell.
For the latter the cell-periodic convention corresponds to a Fourier transform using the real atomic positions, instead of the coordinates of the unit cell.
In this convention the surface spectrum can be continuously deformed to the Wilson loop spectrum \cite{Fid11}.
Here, we only argue for the existence of surface states from the bulk-boundary correspondence, while quantitative statements on the shape of surface states rely on direct calculations, cf. Fig.~3 in the main text.

\begin{figure}
\begin{center}
\includegraphics[width=0.9\textwidth]{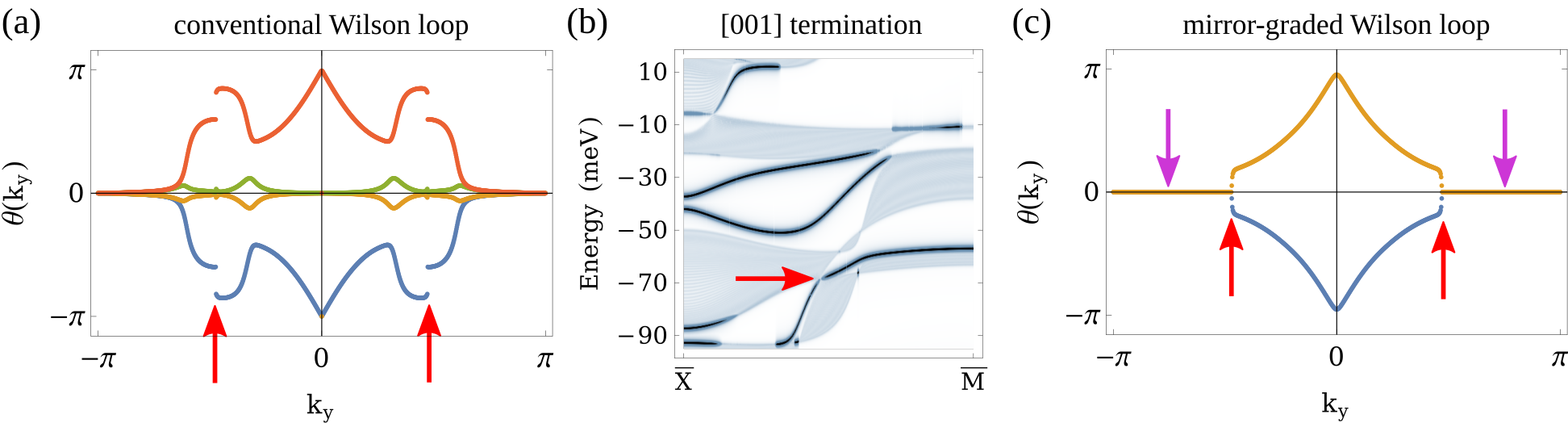}
\caption{Comparison of Wilson loop spectra for the A-type antiferromagnetic phase.
All loops $\mathcal{L}_{k_y}$ are taken at constant $k_x = \pi$ and along the $k_z$ axis.
The red arrows highlight the positions where $k_y$ cuts a Dirac nodal line.
(a) Spectrum of the Wilson loop calculated according to Eq.~(\ref{Eq_fullTBWilsonLoop}) at two filled, degenerate bands.
(b) Corresponding band structure of a slab in [001] termination with highlighted surface states, see also Fig.~3 in the main text.
(c) Spectrum of the mirror-graded Wilson loop calculated according to Eq.~(\ref{Eq_mirrorWilsonLoop}) for two filled bands, where the number of different values of $\theta(k_y)$ at each $k_y$ is half of that in (a), due to the projection on one eigenspace of the glide mirror symmetry $\tilde{m}_x$ in Eq.~(\ref{Eq_mirrorWilsonLoop}).
Purple arrows mark the regions where $\theta (k_y)$ has an exact degeneracy and a flat dispersion with $\theta(k_y) \equiv 0$.
}
\label{fig_ComparisonWilsonLoops}
\end{center}
\end{figure}

We implemented the Wilson loop matrix $\mathcal{W}(\mathcal{L}_{k_y})$ according to Eq.~(\ref{Eq_fullTBWilsonLoop}) for loops $\mathcal{L}_{k_y}$ parametrized by  $(\pi,k_y,k_z)$ for $k_z \in [ - \pi/2, \pi/2 ]$, i.e, $\bm{k}^{(0)} = (\pi,k_y,-\pi/2)$ and $\bm{G} = (0,0,\pi)$, and at all even-integer fillings up to the Fermi energy.
To describe the flow of the Wilson loop spectrum as a function of $k_y$, one considers the complex phases $\theta(k_y)$ of the eigenvalues of the Wilson loop matrix $\mathcal{W}(\mathcal{L}_{k_y})$, see Fig.~\ref{fig_ComparisonWilsonLoops}(a).
We find that these complex phases reach all possible values, i.e., $-\pi$ to $+\pi$, within the Brillouin zone, which due to their relation to the Wannier charge centers implies the presence of surface states, see Fig.~\ref{fig_ComparisonWilsonLoops}(a,b) \cite{Fid11}.
Additionally, in this geometry signatures of the Dirac nodal lines are expected, once $k_y$ is close to the position of the nodal lines.
Indeed, we find jumps in $\theta(k_y)$ once the loop $\mathcal{L}_{k_y}$ passes through one of the nodal lines.
Yet, this approach is not optimal to capture the topology of Dirac nodal lines.
The fourfold degenerate Dirac nodal lines exhibit twice the Berry phase that a twofold degenerate nodal line carries.
In other words, the Berry phase, which is the total of all $\theta(k_y)$ for loops enclosing the nodal line, is trivial.
Thus, we propose in the following a mirror-graded Wilson loop, which makes the topological quantization evident.

To refine the concept of non-Abelian Wilson loops, we use the glide mirror symmetry $\tilde{m}_x$ to block-diagonalize the Hamiltonian before we define the mirror-graded Wilson loop.
We project the Hamiltonian into the occupied subspace belonging to the mirror eigenvalue $+i \mathrm{e}^{i k_z} $ of $\tilde{m}_x$ with the corresponding projector $ \mathcal{P}^+_{\bm{k}} $ and obtain $H^+(\bm{k}) = \mathcal{P}^+_{\bm{k}} H(\bm{k}) \mathcal{P}^+_{\bm{k}} $.
The eigenvalues $\{ \pm i \mathrm{e}^{i k_z} \}$ of $\tilde{m}_x$ exchange over the course of the Brillouin zone, which implies the eigenspace belonging to the eigenvalue $+i \mathrm{e}^{i k_z} $ at $k_z = \pi/2$ has no overlap with the same eigenspace at $k_z =-\pi/2$, whereas the eigenspaces at $k_z =+ \pi$ and $k_z = -\pi$ are isomorphic.
Thus, we propose to extend the loop  $\mathcal{L}$ from the interval $k_z \in [-\pi/2,\pi/2]$ to  $k_z \in  [-\pi,\pi]$.
Analogously to Eq.~(\ref{Eq_fullTBWilsonLoop}) we then define a mirror-graded Wilson loop $\mathcal{W}^+$ using the eigenstates $\left | U^{+,i}_{\bm{k}} \right \rangle $ of $H^+(\bm{k})$,
\begin{align}
	[\mathcal{W}^+(\mathcal{L})]^{mn}
	&=
	\left \langle U^{+,m}_{\bm{k}^{(0)} + 2\bm{G}}  \right |
	W^+(\bm{k}^{(0)} +2 \bm{G} \leftarrow \bm{k}^{(0)})
	\left | U^{+,n}_{\bm{k}^{(0)}}  \right \rangle,
	\label{Eq_mirrorWilsonLoop}
	\\
	W^+(k^{(\gamma)} \leftarrow k^{(\beta)})
	&=
	\prod_{\alpha}^{k^{(\gamma)} \leftarrow k^{(\beta)}}	
	\mathcal{P}^{+,\text{occ}}_{\bm{k}^{(\alpha)}},
	\label{Eq_mirrorWilsonSegment}
	\\
	\mathcal{P}^{+,\text{occ}}_{\bm{k}}
	&=
	\sum^{n_\text{occ}}_{i = 1}
	\left | U^{+,i}_{\bm{k}} \right \rangle
	\left \langle  U^{+,i}_{\bm{k}}  \right |,
\end{align}
where $	\mathcal{P}^{+,\text{occ}}_{\bm{k}} $ projects onto occupied eigenstates of $H(\bm{k})$ with mirror eigenvalue $+i \mathrm{e}^{i k_z} $.
The reciprocal lattice vector to close the loop is now $2\bm{G}$, i.e., for the A-type antiferromagnetic order $2\bm{G} = (0,0,2\pi)$ in units of the inverse lattice constant of SmB$_6$.
We are going to argue that Eq.~(\ref{Eq_mirrorWilsonLoop}) is not missing terms appearing in Eq.~(\ref{Eq_fullTBWilsonLoop}).
Yet, since the order in which the factors are considered differs between Eqs.~(\ref{Eq_fullTBWilsonLoop}) and (\ref{Eq_mirrorWilsonLoop}), the spectrum of the mirror-graded Wilson loop is different.
First, note that in the cell-periodic basis convention the representation of the mirror symmetry comprises a k-\emph{in}dependent matrix and a k-dependent prefactor.
As a result all eigenstates of $\tilde{m}_x$ are constant, which implies in turn that the mirror eigenstates of different eigenvalues at $\bm{k}^{(\alpha+1)}$ and $\bm{k}^{(\alpha)}$ are pair-wise orthogonal.
This allows us to simplify the overlap terms, which occur in the conventional tight-binding Wilson loop, Eq.~(\ref{Eq_fullTBWilsonLoop}), in the form of $\langle U^i_{\bm{k}^{(\alpha+1)}} |U^j_{\bm{k}^{(\alpha)}} \rangle$ for occupied states $i$ and $j$.
This overlap $\langle U^i_{\bm{k}^{(\alpha+1)}} |U^j_{\bm{k}^{(\alpha)}} \rangle$ is only non-zero within each mirror subspace and therefore the projection $ \mathcal{P}^+_{\bm{k}}$ does not remove any non-vanishing terms in the Wilson loop.
But, as mentioned before, the glide-mirror operator eigenvalues exchange and we cannot meaningfully close our Wilson loop at $k_z = \pi/2$.
In other words, the Hamiltonian $H^+(\bm{k})$ projected on the eigenspace of $+i \mathrm{e}^{i k_z} $ evolves into the Hamiltonian $H^-(\bm{k})$ projected on the eigenspace of $-i \mathrm{e}^{i k_z} $, i.e,  $H^+(\bm{k} + (0,0,\pi)) = H^-(\bm{k})$.
In conclusion, Eq.~(\ref{Eq_mirrorWilsonLoop}) captures both eigenspaces of $\tilde{m}_x$, as does Eq.~(\ref{Eq_fullTBWilsonLoop}), but the mirror-graded Wilson loop does so not simultaneously but consecutively in a longer loop.
This leads to Wilson loop matrices $\mathcal{W}^+(\mathcal{L})$ with half the number of rows and columns for the same number of occupied bands.

The mirror-graded Wilson loop, as introduced in Eq.~(\ref{Eq_mirrorWilsonLoop}), remedies the issues of the conventional Wilson loop defined in Eq.~(\ref{Eq_fullTBWilsonLoop}).
We use the new definition and find the Wilson loop spectrum shown in Fig.~\ref{fig_ComparisonWilsonLoops}(c).
A rapid change in the spectrum of the mirror-graded Wilson loop occurs at the position of the nodal line.
It appears that in the vicinity of the Dirac nodal line there is a slope that is too step for the chosen resolution but not a discontinuity as it is seen in Fig.~\ref{fig_ComparisonWilsonLoops}(a).
Notably, at the $k_y$ value where the band structure exhibits a nodal line, the mirror-graded Wilson loop $\theta(k_y)$ has a kink, (red arrow) which separates a dispersive region from a nondispersive flat region (pink arrows).
We will discuss the reason for this behavior in more detail below.
Furthermore, the presence of small avoided crossings in the band structure does not lead to unphysical jumps in the mirror-graded Wilson loop spectrum.
Finally, as far as the plot resolution permits, the Wilson loop spectrum spans the full range of complex phases, see Fig.~\ref{fig_ComparisonWilsonLoops}(c), and thus there is still a connection between the Wilson loop and the surface spectrum  Fig.~\ref{fig_ComparisonWilsonLoops}(b).

To understand the (mirror-graded) Wilson loop spectra, we consider the symmetries that constrain its spectrum.
The sum over the band index $i$ of the complex phases $\theta(k_y)_i$ in the Wilson loop  $\mathcal{W}(\mathcal{L}_{k_y})$ spectrum corresponds to the usual Berry phase, which in the present case is identically zero due to inversion and time-reversal symmetry \cite{Ale14}.
An inversion sends $k_y \to - k_y$ and inverts the direction of the Wilson loop, which implies that the phase fulfills $\theta(k_y)_i = - \theta(-k_y)_j $ for one pair $i$ and $j$ of band indices.
The application of the time reversal operator does not only invert the vector $\bm{k}$ but also adds a complex conjugation, which leads to $\theta(k_y)_i = \theta(-k_y)_i $.
Together inversion and time reversal imply a symmetric spectrum around $\theta =0$ with $\theta(k_y)_i = - \theta(k_y)_j $ and thus a vanishing Berry phase.

After this preliminary discussion of symmetries we can return to the question, why there are flat bands in the spectrum of the mirror-graded Wilson loop in Fig.~\ref{fig_ComparisonWilsonLoops}(c).
We are going to connect the Dirac nodal lines, which are protected by a second mirror symmetry, namely the symmorphic operation $m_z$, and show under which circumstances the mirror-graded non-Abelian Wilson loop is exactly determined by the mirror eigenvalues $\lambda_{m_z} \in \{\pm i\}$ of $m_z$.
While the latter is generally the case for Wilson loops of non-degenerate bands perpendicular to a mirror plane, for degenerate bands this is not the case, cf. Fig.~\ref{fig_ComparisonWilsonLoops}(a).
Nevertheless, the mirror-graded non-Abelian Wilson loop partially recovers this connection between the Berry phases $\theta(k_y)_i$ and the mirror eigenvalues $\lambda_{m_z}$.
The action of $m_z$ on eigenstates $| U^n_{k_z}, \lambda_{\tilde{m}_x} \rangle$ of $\tilde{m}_x$ is given by  $ m_z| U^n_{k_z}, \lambda_{\tilde{m}_x} \rangle \propto | U^n_{-k_z}, -\lambda_{\tilde{m}_x} \rangle$, where the eigenvalue $\lambda_{\tilde{m}_x} $ denotes the mirror subspace used in the mirror-graded Wilson loop, Eq.~(\ref{Eq_mirrorWilsonLoop}).
Here, the $k_x$ and $k_y$ dependence is not shown explicitly.
By using that the mirror $\tilde{m}_x$ eigenvalues and eigenstates are related due to their $k_z$ dependence, one may rewrite $| U^n_{-k_z}, -\lambda_{\tilde{m}_x} \rangle \propto | U^n_{\pi -k_z }, \lambda_{\tilde{m}_x} \rangle$.
Because all appearing proportionality factors have a unit modulus, and thus drop out in the tight-binding Wilson loop, one finds the relation $m_z  W(k^{(\beta)}_z \leftarrow k^{(\alpha)}_z) m_z^\dagger = W(\pi-k^{(\beta)}_z \leftarrow \pi-k^{(\alpha)}_z)  $.
To be specific, this can be obtained by introducing $\mathbbm{1} = m_z^\dagger m_z$ in each inner product contained in $W(k^{(\beta)}_z \leftarrow k^{(\alpha)}_z)$, cf. Eq.~(\ref{Eq_mirrorWilsonSegment}).
In the following we will split the full mirror-graded Wilson loop into two parts, which then can be related by the relation above, yielding
\begin{align}
	[\mathcal{W}^+(\mathcal{L}_{k_y})]^{mn}
	&=
	\left \langle U^{+,m}_{-\pi/2 }  \right |
	W^+(\tfrac{3 \pi}{2} \leftarrow -\tfrac{\pi}{2})
	\left | U^{+,n}_{\pi/2}  \right \rangle
	=
	\left \langle U^{+,m}_{-\pi/2 }  \right |
	W^+(\tfrac{3 \pi}{2} \leftarrow \tfrac{\pi}{2})
	W^+(\tfrac{\pi}{2} \leftarrow -\tfrac{\pi}{2})
	\left | U^{+,n}_{\pi/2}  \right \rangle
	\\
	&=
	\left \langle U^{+,m}_{-\pi/2 }  \right |
	m_z^\dagger \, W^+(-\tfrac{\pi}{2} \leftarrow \tfrac{\pi}{2}) \, m_z
	W^+(\tfrac{\pi}{2} \leftarrow -\tfrac{\pi}{2})
	\left | U^{+,n}_{\pi/2}  \right \rangle
	\\
	&=
	\sum_{opq}
	\left \langle U^{+,m}_{-\pi/2 }  \right |m_z^\dagger  \left | U^{+,o}_{-\pi/2 } \right \rangle
	\left \langle U^{+,o}_{-\pi/2 } \right | W^+(-\tfrac{\pi}{2} \leftarrow \tfrac{\pi}{2})  \left | U^{+,p}_{\pi/2 }  \right \rangle
	\nonumber\\
	&\qquad\quad \times \left \langle U^{+,p}_{\pi/2 }  \right |  m_z  \left | U^{+,q}_{\pi/2 } \right \rangle
	\left \langle U^{+,q}_{\pi/2 } \right | W^+(\tfrac{\pi}{2} \leftarrow -\tfrac{\pi}{2}) \left | U^{+,n}_{\pi/2}  \right \rangle,
	\label{Eq_TBWilsonLoopPlusMirror}
\end{align}
where in the last step identity operators have been introduced that can be directly reduced to summations over occupied bands by the projectors in $W^+(\ldots)$.
In principle the second and fourth term of Eq.~(\ref{Eq_TBWilsonLoopPlusMirror}) cancel if they are multiplied as matrices.
Hence, one finds again the result obtained from inversion and time-reversal symmetry, i.e., $\ln \det \mathcal{W}(\mathcal{L}_{k_y}) = 0$ meaning the Berry phase vanishes.
Furthermore, one may now understand the emergence of flat bands in Fig.~\ref{fig_ComparisonWilsonLoops}(c).
If all occupied bands have the same $m_z$ mirror eigenvalue at $k_z = \pm \pi /2$ one finds that
\begin{align}
	\left \langle U^{+,p}_{\pi/2 }  \right |  m_z  \left | U^{+,q}_{\pi/2 } \right \rangle
	= \lambda_{m_z} \delta_{pq},
	\quad \text{ and thus }\quad
	[\mathcal{W}^+(\mathcal{L}_{k_y})]^{mn}
	= \left \langle U^{+,m}_{-\pi/2 } \right |
	  W^+(-\tfrac{\pi}{2} \leftarrow \tfrac{\pi}{2}) W^+(\tfrac{\pi}{2} \leftarrow -\tfrac{\pi}{2})
	  \left | U^{+,n}_{\pi/2}  \right \rangle
	= \mathbbm{1}^{mn}.
\end{align}
For a Wilson loop equal to the identity matrix all Berry phases are exactly $\theta(k_y)_i = 0$.
Indeed, we see in Fig.~\ref{fig_ComparisonWilsonLoops}(c) that the Wilson loop spectrum is generally dispersive, but is (partially) constant wherever $\lambda_{m_z} $ exchange at a Dirac nodal line.
Note, more complex ways in which Eq.~(\ref{Eq_TBWilsonLoopPlusMirror}) may reduce to a diagonal matrix can also occur.

To summarize our method, it is possible to characterize a Dirac nodal line in the presence of a glide mirror symmetry by introducing a mirror-graded Wilson loop.
We also find that once we consider one eigenspace of the mirror symmetry with the doubled Brillouin zone, the nodal line appears twice and each occurrence is characterized by a $\pi$ Berry phase on a small loop encircling the nodal line.
Nevertheless, the full information about the surface state topology is only found in the flow of the Wilson loop spectrum.
Furthermore, we have shown that the mirror-graded Wilson loop spectrum is not smooth, but instead is expected to exhibit regions with constant eigenvalues in the presence of a second mirror symmetry $m_z$.

\section{Band topology of SmB$_6$} \label{SecTopAnalysis}

In this section we give a comprehensive topological analysis of our proposed SmB$_6$ model, both in the paramagnetic and magnetic phases.
To do so we begin with the already known topological properties of the paramagnetic phase and apply the methodology to our model (see Sec.~\ref{Sec_BandTopPM}). We then analyze the band topology of both the A-type and G-type AFM phases (see Sec.~\ref{Sec_BandTopAafm} and \ref{Sec_BandTopGafm}).
Furthermore, we discuss the properties of the high-field phase, where all moments of SmB$_6$ are aligned (see Sec.~\ref{Sec_BandTopFM}).

Our aim is to work out the key differences in the band topology between the paramagnetic and different magnetic phases of SmB$_6$.
The topological invariants, surface states and possible symmetry protected crossings are considered for the A-type and G-type phases, as well as for the high-field spin-polarized phase.

For the A-type phase we apply the previously introduced method of the mirror-graded Wilson loop (Sec.~\ref{SecTopMethod}).
By the bulk-boundary correspondence, this relates the topology of the nodal lines in the bulk to the drumhead states at the surface.
Since the formation of a magnetic order gives rise to a significant reorganization of the electronic bands, we begin each discussion with an analysis of the spatial symmetries and the back folding of the bands.
This is a prerequisite for the subsequent discussion of band structures and topological invariants.

As the details of the magnetic structure are not yet experimentally determined, we have analyzed our proposed model for the possible magnetic structures with ordering vector $q=(\pi,0,0)$ and $q=(\pi,\pi,\pi)$, corresponding to A- and G-type orderings, respectively. We introduce the magnetic order to our model $\mathcal{H}$, Eq.~(\ref{eq:H}), by appropriately enlarging the unit cell and including the exchange Hamiltonian as Zeeman terms on a mean-field level. For the antiferromagnetic phases, we consider the valence state with $v=2.74$, i.e. above a critical value of $v_c=2.7$ for the onset of AFM order.  We use the model parameters as specified in the main text and Sec.~\ref{SecTBModel} above.
 	
\subsection{Paramagnetic phase}
\label{Sec_BandTopPM}

\subsubsection{Band structure of the paramagnetic phase}

First, we consider the paramagnetic state with average valence  $v=2.56$, corresponding to ambient pressure.
SmB$_6$ crystallizes in the cubic space group $Pm\bar{3}m$ (No. 221).
Due to the presence of both inversion and time-reversal symmetry all bands are twofold degenerate.
From the fact that this space group contains only symmorphic operations, it follows that accidental symmetry protected band crossings can only occur along fourfold and threefold rotation axes.
On all other positions in k space the bands exhibit at least a small direct gap.
As seen in Fig.~\ref{fig:s1} there are movable Dirac points along $\Gamma$-X.
The irreducible representations of the double groups at the time-reversal invariant momenta (TRIMs) $\Gamma$ and R are four dimensional, which implies fourfold degeneracies (including spin) \cite{Elc12}.
These Dirac points are slightly below the Fermi energy for the bands with f-orbital character and more than one 1~eV above the Fermi energy for the d bands.
The former type of Dirac points is exceptionally massive, compared to well known examples like graphene \cite{Nov05} or Cd$_3$As$_2$ \cite{Neu12}.
Note that the adjacent bands above and below the Dirac points are completely filled, thus no unusual topological transport is expected from these crossings.

\subsubsection{Topological invariants of the paramagnetic phase}
\label{Sec_TopInvariantsPM}

Due to the full band gap and time-reversal symmetry, SmB$_6$ can be considered as an insulator in the symmetry class AII of the Altland-Zirnbauer classification.
By employing the inversion eigenvalues $\xi_n(\Gamma_i) = \pm 1$ of all the bands labeled by $n$ at the TRIMs $\Gamma_i \in\{ \Gamma,\text{X}, \text{M}, \text{R} \}$ we can assess the strong (weak) topological $\mathbb{Z}_2$ invariant $\nu_0$ ($\nu_1,\nu_2,\nu_3$) according to Ref.~\cite{Fu07} by
\begin{align}\label{Eq_definitionZ2}
(-1)^{\nu_j} = \prod_{i \in \textrm{TRIM}} \delta_i,
\qquad\text{where}
\qquad
\delta_i = \prod_{m=1}^N \xi_{2m}(\Gamma_i),
\end{align}
and $i$ labels the TRIMs of the system.
The second product in Eq.~(\ref{Eq_definitionZ2}) is over the $N$ occupied Kramers pairs of bands, where we note that for a given Kramers pair $\xi_{2m-1} ( \Gamma_i ) = \xi_{2m} ( \Gamma_i )$.
For the calculation of $\nu_0$ all eight TRIMs must be used for $\Gamma_i$ in Eq.~(\ref{Eq_definitionZ2}), whereas for the weak invariants $\nu_1, \nu_2,$ or $\nu_3$ only the TRIMs on the planes $k_x = \pi, k_y  =\pi,$ or $k_z = \pi$ are considered, respectively.
For both the weak and strong invariants the presence of spatial symmetries and band degeneracies simplifies the calculation.
The Dirac points at $\Gamma$ and R contain only bands with the same inversion eigenvalue.
Thus, for an even number of filled bands (such that $\delta_i$ is well-defined at each TRIM) the respective values of $\delta_i$ are $\delta_\Gamma = \delta_\mathrm{R}$.
This applies to our model of SmB$_6$, because for the above mentioned parameters two bands are occupied, and the Fermi level lies within a band gap.
Due to the threefold rotation the three equivalent TRIMs corresponding to X carry the same inversion eigenvalues  $\xi_n(\Gamma_i)$.
The same holds for M.
Thus, for the strong topological invariant $\nu_0$ two out of the three TRIMs X and M cancel cancel in the product of $\delta_i$.
Together with the trivial contributions from the Dirac points at $\Gamma$ and R the resulting strong $\mathbb{Z}_2$ invariant is given by $(-1)^{\nu_0} = \delta_\mathrm{X} \delta_\mathrm{M}$.
As expected, we obtain from our tight-binding model at half filling $\nu = 1$ due to $\delta_\mathrm{X} = -1$ and $\delta_\mathrm{M} = 1$.
In the presence of a threefold rotation around the [111] axis the weak invariants fulfill $\nu_1 = \nu_2 = \nu_3 $ and are given by $(-1)^{\nu_3} =  \delta_\mathrm{X}$, because in space group  $Pm\bar{3}m$ the fourfold representation at R implies $ \delta_\mathrm{R} = 1$ independently of the material.
Thus, at half-filling the weak invariants  $\nu_1 = \nu_2 = \nu_3= 1$ are nontrivial, see Table~\ref{PM_TopInvariants}.

Due to the presence of mirror symmetries we can also define mirror Chern numbers $\mathcal{C}_{m_z, k_z =0}$ and $\mathcal{C}_{m_z, k_z = \pi}$ within the mirror planes $k_z = 0$ and $k_z = \pi$, respectively \cite{Kim18}.
These mirror Chern numbers are defined as $\mathcal{C}_{m_z, k_z} = \mathcal{C}_{+} -  \mathcal{C}_{-}$, where $ \mathcal{C}_{\pm}$ is the Chern number calculated at a fixed value of $k_z$ within the subspace of states belonging to one mirror eigenvalue $\pm i$, see Table~\ref{PM_TopInvariants}.
Half of the mirror Chern number $\mathcal{C}_{m_z, k_z}/2$ is equal to the number of surface states that cross the bulk gap in the same direction.
Hence, the subsystem with $k_z = 0$ exhibits two surface states, whereas for $k_z = \pi$ there is one surface state.
This is consistent with the previously calculated $\mathbb{Z}_2$ invariants, which imply an odd (even) number of surface Dirac cones for $k_z = \pi$ ($k_z = 0$), cf. Fig.~3(g) in the main text.
Note that in the presence of time-reversal symmetry the regular Chern number $\mathcal{C} = \mathcal{C}_{+} +  \mathcal{C}_{-}$ vanishes.
We list all the discussed topological invariants for paramagnetic SmB$_6$ in Table~\ref{PM_TopInvariants}.

\begin{table}
\begin{tabular}{c|c|c|c|c||c|c||c|c}
Band $N$ & $\delta_\Gamma$ & $\delta_\text{X}$ &  $\delta_\text{M}$ &  $\delta_\text{R}$ & $\nu_3$ & $\nu_0$ & $\mathcal{C}_{m_z, k_z = 0}$ & $\mathcal{C}_{m_z, k_z = \pi}$ \\
\hline \hline
1 & -- & 1 & -1 & -- & --  & --  &-- &-- \\
\hline
2 & 1 & -1 & 1 & 1 & 1 & 1 & 4 & 2 \\
\hline
3 & -- & 1 & 1 & -- & -- & -- & -- &-- \\
\hline
4 & 1 & 1 & 1 & 1 & 0  & 0 & 0 & 0
\end{tabular}
\caption{\label{PM_TopInvariants}
Topological invariants of the paramagnetic phase.
The product of all inversion eigenvalues up to the number of occupied bands (first column) are given for each TRIM in the second to the fifth column.
The last four columns list the topological invariants, which characterize the occupied bands, comprising the strong $\mathbb{Z}_2$ invariant $\nu_0$, the weak $\mathbb{Z}_2$ invariant $\nu_1=\nu_2 =\nu_3$, as well as the mirror Chern numbers $\mathcal{C}_{m_z k_z = 0,\pi}$.
}
\end{table}

\subsection{A-type antiferromagnetic phase}
\label{Sec_BandTopAafm}

\subsubsection{Band structure of the A-type antiferromagnetic phase}

In the A-type antiferromagnetic phase the regular time-reversal symmetry $T$ is broken, but there exists a modified time-reversal operation $T_A = T t(0,0,1)$, which comprises the regular time reversal $T$ and a translation $t(0,0,1)$.
A fixed orientation of the magnetic moments in the A-type phase removes the threefold rotation symmetry of the paramagnetic phase.
The corresponding magnetic space group is $P_{2c}4/mm'm'$ (No. 123.15.1013 in the OG convention).
Accidental Dirac points occur only along one of the two fourfold rotation axes, i.e., on $\Gamma$-Z but not M-A, just as in the paramagnetic (and G-type) phases \cite{elc2020}.
As discussed in Sec.~\ref{SecTopMethod}, the combination of inversion and time-reversal $T_A$ leads to a twofold degeneracy of all bands.

As described in the main text and different from the G-type case there are two different band representations on the mirror plane $k_z = \tfrac{\pi}{2}$, where different mirror eigenvalues are paired by the combination of time reversal $T_A$ and inversion $P$.
This means SmB$_6$ in its A-type phase does not only host Dirac points on a fourfold rotation axis, but also Dirac nodal lines that are absent in the G-type phase, see Fig.~\ref{fig:AtypeBands} and Fig.~3 in the main text.
Dirac points appear between bands (2,~3) and (4,~5) at the energies -62~meV and -37~meV,  respectively, below the Fermi level.

Additionally, there are accidental but symmetry-protected nodal lines around the A point between the bands (2,~3) and close to the Z and R points between the bands (4,~5).
The latter nodal lines are close to the Fermi energy and are partially of type II.
These type-II nodal lines exhibit tilted bands along some radial cuts of the band structure intersecting the nodal line, while in the direction orthogonal to both the mirror plane and the nodal line, i.e., along $k_z$ they exhibit regular cone-like dispersions \cite{He18}.
The nodal lines near R are weakly dispersive spreading from -6~meV to 5~meV in energy.
Near Z the nodal lines are effectively dispersionless, covering energies from -32~meV to -31~meV. Both types of nodal lines have a diameter of up to one fifth of the Brillouin zone (see Fig. 3(c) of the main text).

\subsubsection{Topological invariants of the A-type antiferromagnetic phase}

\begin{figure}
\begin{center}
\includegraphics[width=10cm]{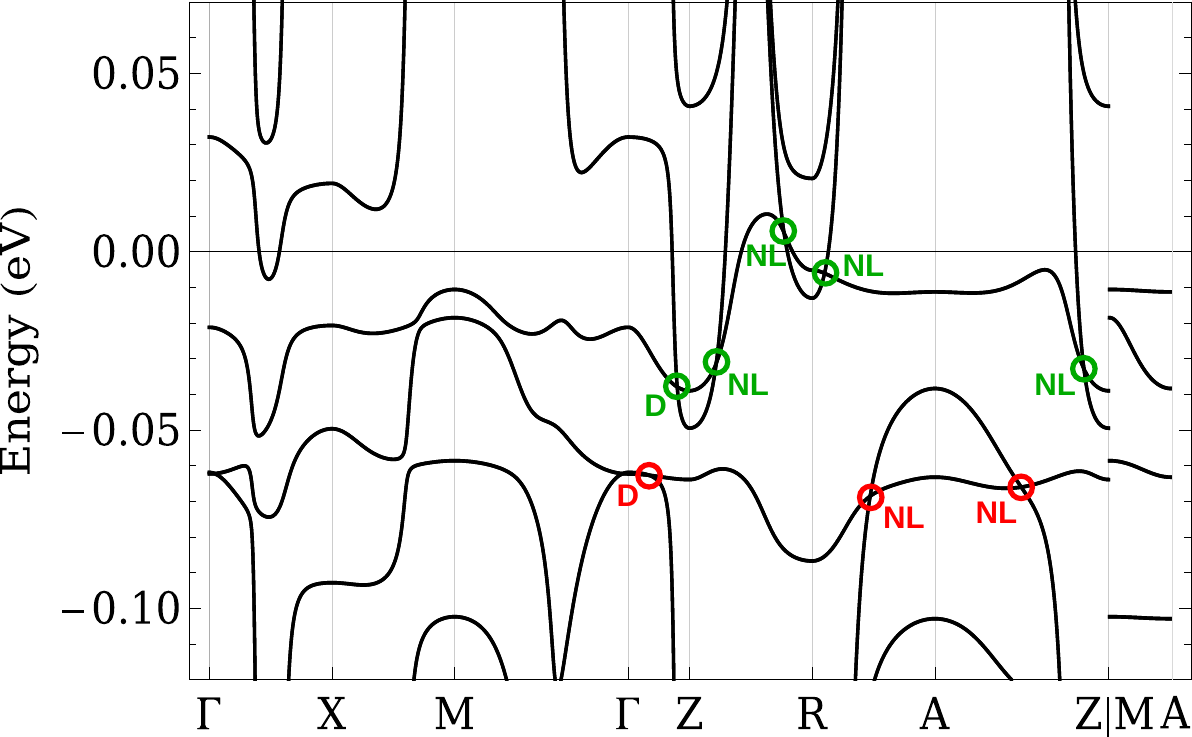}
\caption{Bandstructure of SmB$_6$ in the A-type antiferromagnetic phase.
The circles indicate Dirac points~(D) and nodal lines (NL), where the colors red and green indicate crossings between bands (2,3) and (4,5), respectively, as in Fig. 3 of the main text.}
\label{fig:AtypeBands}
\end{center}
\end{figure}

We group our discussion of the topological invariants according to the associated surface states that appear on the [010] and [001] surfaces, as discussed in the main text.

\paragraph*{Topology of [010] surface states.--} To assess the topology relevant for the [010] termination we consider two invariants that can be defined for the $k_z =0$ plane.
We note that for SmB$_6$ in the A-type AFM phase there exists a full gap between the fourth and the fifth band in the $k_z = 0$ plane, see Fig.~\ref{fig:AtypeBands}.
Hence, for this filling we can define in the $k_z = 0$ plane a weak $\mathbb{Z}_2$ invariant $\nu_3$ as well as a mirror Chern number $C_{m_z}$.
First we note that the gapped plane $k_z = 0$ can be understood as an antiferromagnetic topological insulator characterized by the weak $\mathbb{Z}_2$ invariant $\nu_3$ \cite{Mon10}.
This is possible because within the $k_z = 0$ plane the time-reversal operation $T_A = T t(0,0,1)$ behaves identically to the regular time reversal $T$.
As for the paramagnetic phase, Sec.~\ref{Sec_TopInvariantsPM},  we determine $\nu_3$ from the inversion eigenvalues \cite{Fu07}.
Assuming that the lowest four bands are filled, we find $\nu_3 =1$ indicating a topological gap in the $k_z = 0$ plane.
This comes as no surprise, since the paramagnetic phase of SmB$_6$ is also nontrivial with $\nu_3 =1$, cf. Table~\ref{PM_TopInvariants}.
During the magnetic transition between these two phases the inversion eigenvalues of the plane with $k_z = \pi$ are folded back to the $k_z = 0$ plane, and thus every second band gap within the $k_z=0$ plane is topologically nontrivial in the A-type AFM phase \cite{Mon10}.
In addition, it turns out that also for the odd band gaps (i.e., for an odd number of filled bands), the weak invariant $\nu_3$ is nontrivial, with the exception of the gap between band 5 and 6, see in Table~\ref{Atype_TopInvariants}.
Note, that the contribution of the TRIM X to $\nu_3$ cancels since it occurs in two copies related by the fourfold symmetry.
Furthermore, we remark that the Kramers pairs at the TRIMs within the $k_z = \pi$ plane (i.e., at Z, R and A) consist of two states with different inversion eigenvalues, as a result of the time-reversal operation $T_A$.
As a consequence, the invariants $\nu_i$ corresponding to the planes $k_x = 0, \pi$, $k_y=0, \pi$, and $k_z = \pi$ are ill-defined, which agrees with the deviation from the  AII topological classification due to $T_A^2 = +1$ at $k_z = \pi$.

\begin{table}
\begin{tabular}{c|c|c|c||c||c}
Band $N$ & $\Gamma$ & X & M  & $\nu_3$ & $\mathcal{C}_m$ \\
\hline \hline
1 & 1 & 1 & -1 & 1 & 6 \\
\hline
2 & -1 & -1 & 1  & 1 & 6 \\
\hline
3 & 1 & 1 & -1  & 1 & 2 \\
\hline
4 & -1 & -1 & 1  & 1 & 6 \\
\hline
5 & 1 & 1 & 1  & 0 & 4
\end{tabular}
\caption{\label{Atype_TopInvariants}
Topological invariants of the A-type antiferromagnetic phase.
Inversion eigenvalues at the TRIMs are given in columns 2, 3, and 4, for the different band fillings.
$\nu_3$ denotes the weak $\mathbb{Z}_2$ invariant associated to the plane $k_z = 0$.
In the last column the mirror Chern numbers $\mathcal{C}_m$ are given for the $k_z = 0$ plane for fillings up to the band given in the first column.
}
\end{table}

Second, we note that within the mirror plane $k_z=0$  the mirror Chern number $\mathcal{C}_{m_z}$ can be defined, since within this plane $PT_A$ pairs different $m_z$ mirror eigenvalues.
The values of the mirror Chern number $\mathcal{C}_{m_z} = \mathcal{C}_+ - \mathcal{C}_- $ are given in Table~\ref{Atype_TopInvariants}.
We note that for all considered band fillings the mirror Chern number is nonzero, while the total Chern number $\mathcal{C}=\mathcal{C}_+ + \mathcal{C}_-$ is always zero.
Bands with a non-zero value of the mirror Chern number are expected to exhibit surface states for lattice terminations that do not break the mirror symmetry $m_z$, e.g., the [010] termination.
The two topological invariants, $\nu_3$ and $\mathcal{C}_{m_z}$, both describe the $k_z = 0$ plane and are consistent.
If $\mathcal{C}_{m_z}/2$ is odd and there must be an odd number of surface states that cross the bulk band gap, then the weak $\mathbb{Z}_2$ invariant is always nontrivial, i.e. $\nu_3 = 1$.
For band 5 the mirror Chern number predicts an even number of surface states, which is in agreement with $\nu_3 = 0$.

The surface states for the [010] termination as function of energy and surface momentum $k_x$ are shown in Fig. 3(g) of the main text.
We observe a non-trivial surface state connecting bulk and valence bands with a different number of surface Dirac cones at $\bar{\Gamma}$ than at $\bar{X}$, in full agreement with the weak $\mathbb{Z}_2$ invariant $\nu_3 =1$ (see Table~\ref{Atype_TopInvariants} and the  black arrows in Fig. 3(g) of the main text).
Within the gap between bands 4 and 5, we observe three surface states, in agreement with the mirror Chern number $\mathcal{C}_{m_z}=6$.
We note that a nontrivial $\mathbb{Z}_2$ invariant has also been found for a related model in Ref.~\cite{Kim18}, whereas we obtain larger values of the mirror Chern number enabled by the increased number of bands stemming from the relevant $e_g$ and $\Gamma_8$ quartet.
The mirror Chern numbers of the paramagnetic phase at half filling add up to the mirror Chern number of the A-type phase of the $k_z = 0$ plane, which corresponds to how the $\mathbb{Z}_2$ invariants of the paramagnetic and antiferromagnetic phase have been found to be related above.
This concludes our analysis of the topological invariants relevant to the [010] surface termination.

\begin{figure}
\begin{center}
 \includegraphics[width=14.5cm]{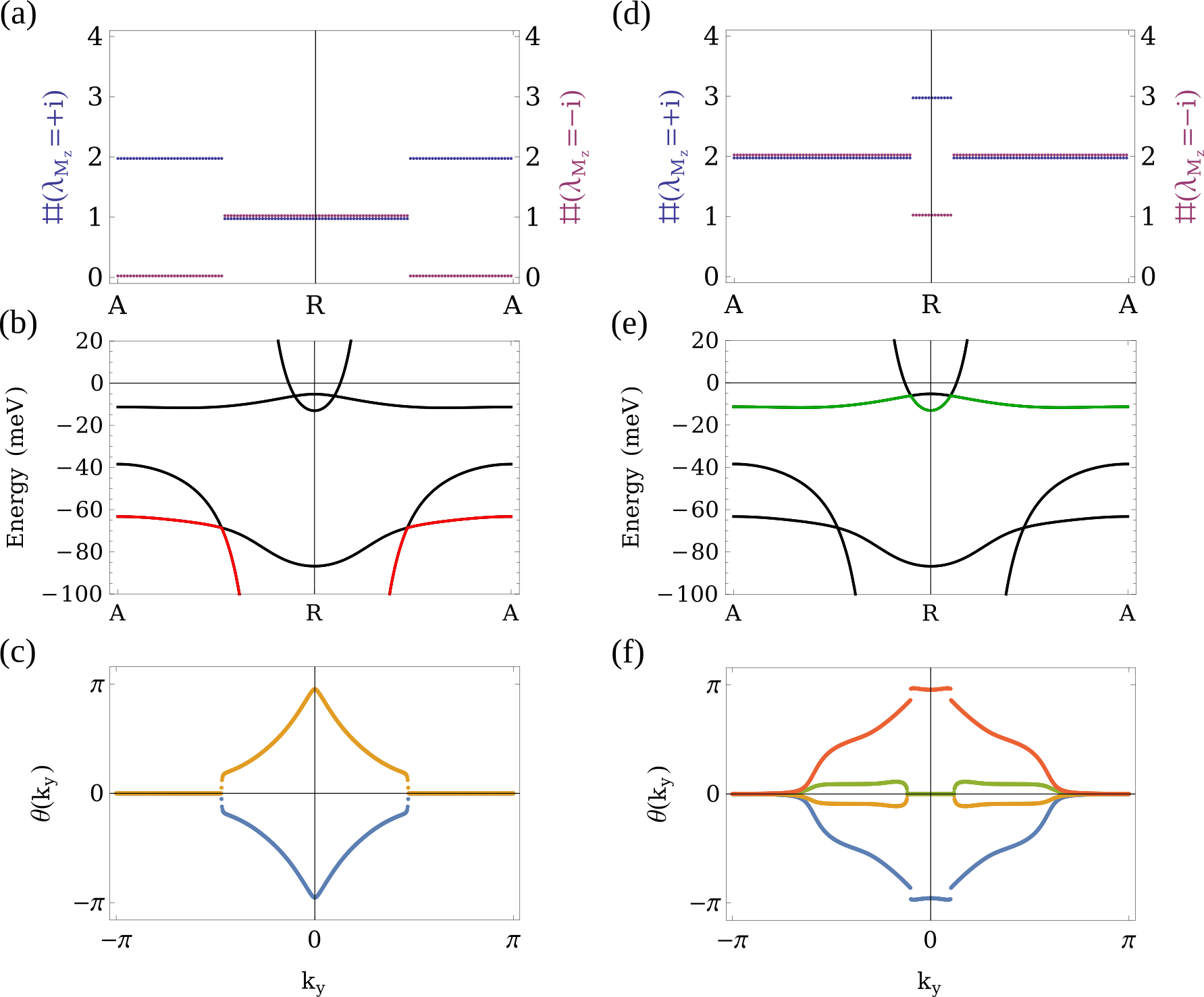}
\caption{Relation of mirror eigenvalues, nodal lines, and the spectral flow of the Wilson matrix in the A-type antiferromagnetic phase.
(a-c) two occupied bands, (d-f) four occupied bands.
(a,d) Multiplicity of mirror eigenvalues for the occupied bands not including the twofold degeneracy due to the  $PT_A$ symmetry.
(b,e) Band structure, the considered band is highlighted with red (green) corresponding to band 2 (4).
(c,f) Phases of the mirror-graded Wilson loop spectra as a function of $k_y$, for loops along the $k_z$ direction with fixed $k_x = \pi$.
}
\label{fig:AtypeMirrorWilsonBands}
\end{center}
\end{figure}

\paragraph*{Topology of [001] surface states.--} Next, we consider the bulk topology relevant for the [001] termination.
First, we consider the distinguishing feature of the A-type magnetic order, i.e., the emergence of mirror symmetry protected nodal lines.
The nodal lines on the plane $k_z = \pi/2$ have a codimension $p = d - d_{FS} = 2$.
Further, time reversal $T_A$ anticommutes with the mirror symmetry $m_z$, while squaring to $T_A^2 = +1$.
This classifies these nodal lines in the presence of such a mirror symmetry $m_z$ as class AI, and they are thus trivial for $p=2$ \cite{Chi14}.
However, we note that in the cited classification scheme the present nodal line shape has not been considered \cite{Chi14}.
Thus, a closer consideration is needed.
The mirror symmetry $m_z$ protects the nodal lines and its eigenvalues $\lambda_{m_z}$ can be used to label the bands.
The values of the mirror eigenvalues $\lambda_{m_z}$ are shown in Fig.~\ref{fig:AtypeMirrorWilsonBands}(a,d).
At the nodal lines mirror eigenvalues cross, which is reflected in the mirror-graded Wilson loop spectrum, see below.

\begin{figure}
\begin{center}
\includegraphics[width=6.5cm]{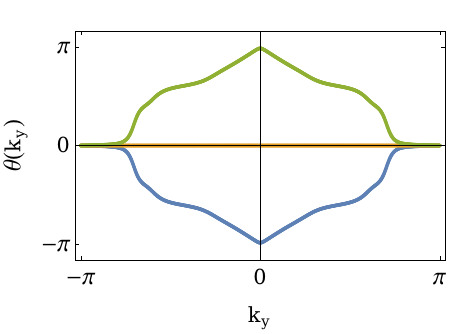}
\caption{\label{fig:AtypeWilsonNoNL}
Mirror-graded Wilson loop spectrum for three occupied bands in the proximity of an avoided crossing.
Wilson loop along the $k_z$ direction at $k_x = \pi$ as function of $k_y$.
The rapid change in the spectrum coincides with a weakly gapped bulk band crossing, cf. surfaces states in Fig.~3(e) (blue arrow) of the main text.
}
\end{center}
\end{figure}

We find that the surface states visible in Fig.~3(e) in the main text can be inferred from the Wilson loop spectrum with periodic boundary conditions.
The Wilson loop spectrum is calculated on mirror planes left  invariant by $\tilde{m}_x$ and $\tilde{m}_y$, which are glide mirror symmetries with a translation along the [001] direction.
In the matrix block of one mirror eigenvalue $H^+(\bm{k})$ all nodal lines are singly degenerate and thus it is possible to determine a Berry phase on a small loop enclosing each nodal line.
The result is a $\pi$ Berry phase for each of the nodal lines.
Yet, this is insufficient to make the actual connection to the surface states, because a quantization of the Berry phase leaves a dependence on the chosen surface termination.

To reach a comprehensive description we thus employ the methodology introduced in Sec.~\ref{SecTopMethod} and determine the Wilson loop spectrum for loops along the $k_z$ direction at fixed $k_x$ and variable $k_y$.
In Fig.~\ref{fig:AtypeMirrorWilsonBands} the analysis is exemplarily shown for the $k_x = \pi$ plane, which determines the surface state along the line $\overline{\text{X}}$-$\overline{\text{M}}$, see Fig.~3(e) in the main text.

For two occupied bands, see Fig.~\ref{fig:AtypeMirrorWilsonBands}(a-c) there is an extended region, where all  bands are described by a single mirror eigenvalue, $\lambda_{m_z} = +i$.
As a result the Wilson loop spectrum is pinned to $\theta = 0$ as shown before in Sec.~\ref{SecTopMethod}.
Starting at the nodal line there is a rapid change in the Wilson loop spectrum, which is accompanied by a surface state in the gap between the second and third band as discussed in the main text and Fig.~3(e) (red arrow).

For four occupied bands, see Fig.~\ref{fig:AtypeMirrorWilsonBands}(d-f), bands of both mirror eigenvalues are always occupied.
Nevertheless, on one side of the nodal line a constant Wilson loop eigenvalue with the phase $\theta = \pi$ emerges and the flow of eigenvalues corresponds to the presence of surface states, as seen in  Fig.~3(e) (green arrow) of the main text.
To support the bulk origin of these surface states, we have confirmed that a featureless potential term applied to the surface of the open system does not remove the surface states.

Finally, a comment is in place regarding the prominent surface state between the third and fourth band on the line $\overline{\text{X}}$-$\overline{\text{M}}$, see Fig.~3(e) (blue arrow) of the main text, which appears to end at an avoided crossing of the bulk spectrum, see Fig.~\ref{fig:AtypeBands}.
Note, the crossing is avoided because on the $k_z = 0$ plane no symmetry protects band crossings.
Although the crossing is gapped, it exhibits smooth features in the Wilson loop spectrum, which are similar to those at nodal lines, see Fig.~\ref{fig:AtypeWilsonNoNL}, i.e., there are nearly flat bands within a region demarcated by the nodal line.
The existence of this feature can be understood from the backfolding of bands for the magnetic unit cell, but unlike for the nodal lines the mirror symmetry $m_z$ is insufficient to protect the crossing.
There is also no $\mathbb{Z}_2$ invariant that could be associated to these gapped Dirac surface states, because on the plane $k_x = \pi$ the relation $T_{\text{A}}^2 = -1$ does generally not hold.

\subsection{G-type antiferromagnetic phase}
\label{Sec_BandTopGafm}

\subsubsection{Band structure of the G-type antiferromagnetic phase}

\begin{figure}
\begin{center}
\includegraphics[width=10cm]{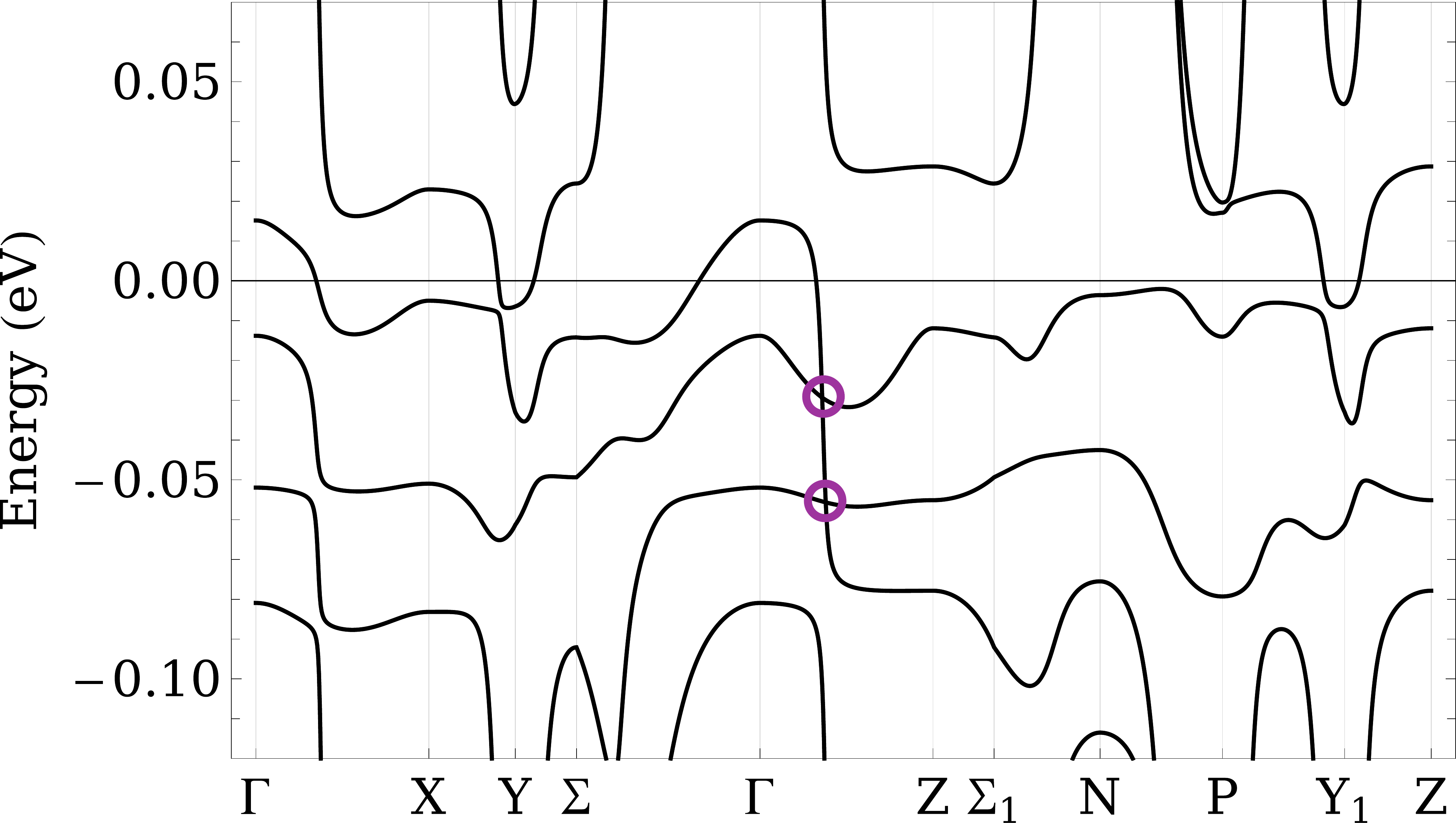}
\caption{Bandstructure of SmB$_6$ in the G-type antiferromagnetic phase. The purple circles highlight the Dirac points that are shown in Fig.~3(d) of the main text.}
\label{fig:GtypeBands}
\end{center}
\end{figure}

The onset of the magnetic order increases the unit cell size and reduces the time-reversal symmetry to the operation $T_G = T t(1,0,0)$ with $T = i \sigma_y K$ and the complex conjugation $K$.
This corresponds to the space group $P_I4/mm'm'$ (No. 123.19.1017 in the OG convention).
Since the threefold rotation symmetry is broken by the magnetic order, crossings of higher degeneracies may only occur along the remaining fourfold rotation axis $\Gamma$-Z, see Fig.~\ref{fig:GtypeBands}.
Dirac points at TRIMs present in the paramagnetic phase are gapped in the magnetic phase \cite{elc2020}.
Notably, unlike for the A-type case, there is no band crossing near the Fermi level for the G-type antiferromagnetic phase.

\subsubsection{Topological invariants of G-type antiferromagnetic phase}

The Brillouin zone geometry of the G-type magnetic order allows for the definition of three weak $\mathbb{Z}_2$ invariants $\nu_{1},\nu_{2},$ and $\nu_{3}$, even though time-reversal symmetry is broken.
This is different to the A-type antiferromagnetic phase, where only $\nu_3$ can be defined.
Interestingly, while each of the weak invariants is determined on a different plane defined by $k_{x} = 0$, $k_{y} = 0$, or $k_{z} = 0$, for all three indices the inversion eigenvalue formula uses the same TRIMs.
The latter statement is evident from the body-centered Brillouin zone shown in Fig.~3(d) of the main text, where all mirror planes contain the same points, i.e., $\Gamma$, X and Z.
Nevertheless, the nonsymmorphic nature of the time-reversal symmetry prevents the definition of the strong invariant $\nu$, because only at $\Gamma$, X and Z the necessary condition $T_G^2 = -1$ holds, while at the TRIM N one finds $T_G^2 = +1$.
The inversion eigenvalues for the former TRIMs are listed in Table~\ref{Gtype_TopInvariants}.
The weak invariants $\nu_i$ can be determined using $(-1)^{\nu_i} = \delta_\Gamma \delta_Z$, where the point X drops out due to its even multiplicity.
Thus, one finds that for fillings 1, 2, 3, and 4, the weak invariants are nontrivial, while for filling 5 they are trivial.
In the surface band structure for the G-type case the bands overlap, see Fig.~3(f,h) in the main text.
Nevertheless, for the [001] termination in the vicinity of the Fermi energy the overlap of bands is sufficiently small so that one can confirm the correspondence between the invariant and the surface spectrum.

\begin{table}
\begin{tabular}{c|c|c|c||c||c}
Band $N$  & $\Gamma$ & X & Z & $\nu_1= \nu_2 = \nu_3$  & $\mathcal{C}_m$ \\
\hline \hline
1 & -1 & 1  & 1 & 1 & 6 \\
\hline
2 & 1 & -1 & -1 & 1 & 10 \\
\hline
3 & -1 & 1 & 1 & 1 & 2 \\
\hline
4 & 1 & -1 & -1 & 1 & 6 \\
\hline
5 &1 & 1 & 1 & 0 & 4
\end{tabular}
\caption{\label{Gtype_TopInvariants}
Topological invariants of the G-type antiferromagnetic phase.
Inversion eigenvalues at the TRIMs are given from the second to the fourth  column taking into account all bands up to the one labeled in the first column.
$\nu_{1},\nu_{2},\nu_{3}$ denote the weak $\mathbb{Z}_2$ invariant associated to the plane $k_x = 0, k_y = 0$ and $k_z = 0$, respectively.
In the last column the mirror Chern numbers $\mathcal{C}_m$ are given for the $k_z = 0$ plane.
}
\end{table}

As before we refine this analysis using crystalline topological invariants.
We will use the symmorphic mirror symmetry $m_z$ and that the Hamiltonian is gapped on the $k_z = 0$ mirror plane.
The mirror Chern number $\mathcal{C}_m$ is consistent with the weak invariant $\nu_3$, see Table~\ref{Gtype_TopInvariants},  and it is at the Fermi energy identical to the A-type phase.
If only two (or three) bands are occupied and the fourfold rotation symmetry is unbroken, then there are two Dirac points on the axis $\Gamma$-Z, hence a mirror Chern number is not well-defined on the planes $k_{x} = 0$ and $k_{y} = 0$.
While the same is technically true for the $\mathbb{Z}_2$ invariant, we can already determine their values from inversion eigenvalues assuming weakly gapped Dirac nodes.
The latter is not possible for the A-type antiferromagnetic phase, where the relevant mirror planes, $k_{x} = 0$ and $k_{y} = 0$, fall into a trivial topological class even if the bands would be gapless.

\begin{figure}
\begin{center}
\includegraphics[width=14.5cm]{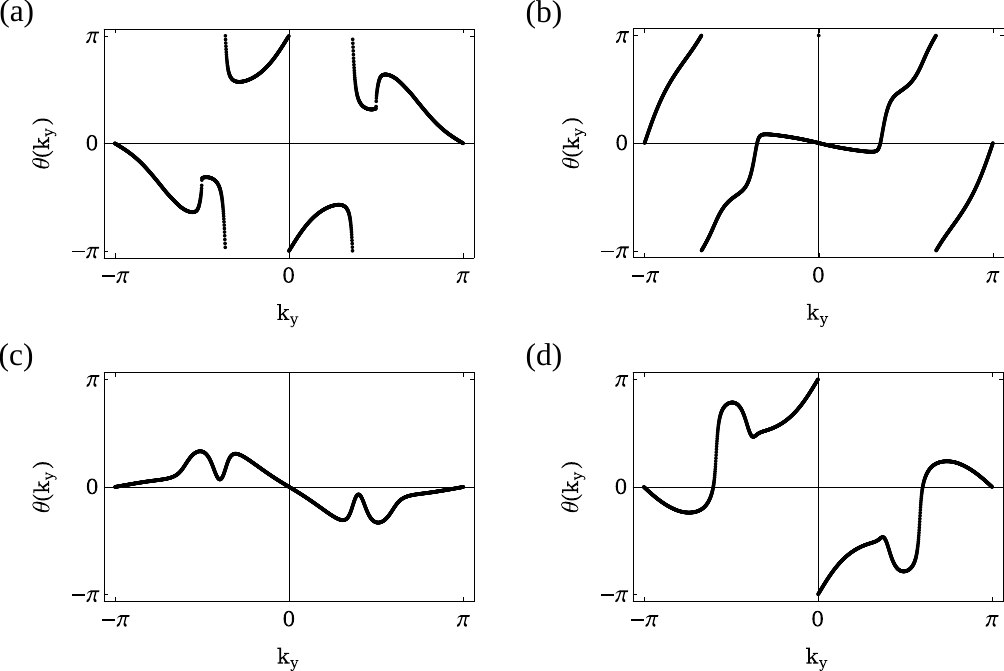}
\caption{Mirror Berry phase winding of SmB$_6$ in the G-type antiferromagnetic phase.
(a)-(d) correspond to the case of 1 to 4 occupied bands, respectively.
The non-Abelian mirror Berry phase $\theta(k_y)$ is computed as a function of $k_y$, for closed loops in the $k_z$-direction at constant $k_x = 0$.
The mirror Berry phases wind once (a,d) or twice (b).
At $k_y = 0$  (b) and (c) are only well-defined in the absence of the Dirac points on the $\Gamma$-Z axis.
}
\label{fig:GtypeBerryPhases}
\end{center}
\end{figure}

To further analyze the surface states at $k_{x,y} = 0$, let us consider the spectral flow of Wilson loop eigenvalues in the subspace of one eigenvalue of mirror symmetry.
Due to the Brillouin zone's extension along $k_z$ the technically nonsymmorphic glide mirror symmetries $\tilde{m}_x$ and $\tilde{m}_y$ do not lead to an exchange of the mirror eigenvalues when moving from $k_z = -\pi$ to $k_z = +\pi$.
Hence, we do not need to use our mirror-graded Wilson loop  formalism, but can just consider a regular mirror Berry phase, which is the sum of all phases of Wilson loop eigenvalues.
The non-Abelian mirror Berry phases for loops along the $k_z$ axis within the $k_x = 0$ plane as function of $k_y$ are given in Fig.~\ref{fig:GtypeBerryPhases}.
We find a winding corresponding to a mirror Chern number of $\mathcal{C}_m^x = -1,+2,0,+1$ for the case of one up to four occupied bands.
While the mirror Berry phase is continuous in Fig.~\ref{fig:GtypeBerryPhases}, this is not necessarily true if we assume a different orientation of loops within the same mirror plane.
By taking loops along $k_y$ with a mirror Berry phase as function of $k_z$, one finds jumps in the Berry phase at the positions of the Dirac points.
If the fourfold symmetry is broken by some perturbation, then these jumps become smooth and the mirror Chern numbers take the values we have given above independently of the orientation of Wilson loops.

To summarize the analysis of the G-type order, we find that unlike for the A-type phase all weak invariants are well-defined and instead of nodal line surface states the [001] termination exhibits a surface Dirac cone, see the black arrow at $E = 18$~meV in Fig. 3(f).

\subsection{High-field limit}
\label{Sec_BandTopFM}

\subsubsection{Band structure in the high-field limit}

\begin{figure}
\begin{center}
\includegraphics[width=10cm]{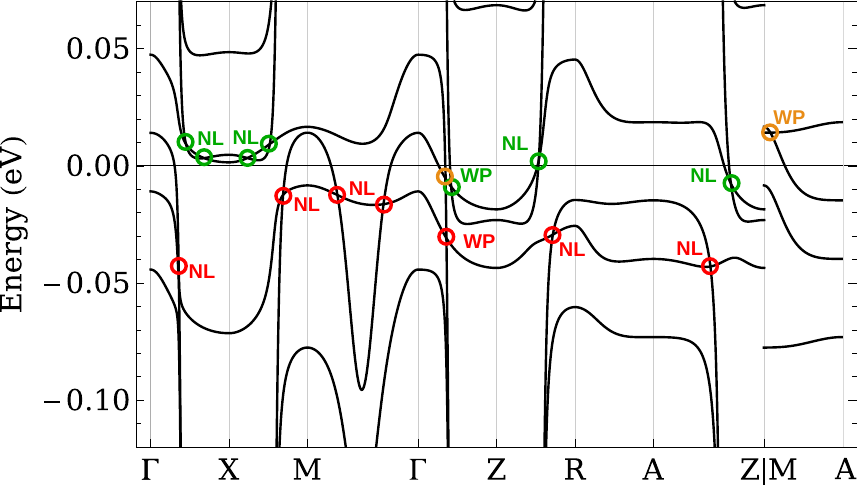}
\caption{Bandstructure of SmB$_6$ with valence 2.73 assuming moments aligned along z direction. The red, orange and green circles highlight the Weyl points and nodal lines.
The red, orange and green color corresponds to band crossings between bands (2,3), (3,4), and (4,5), respectively. }
\label{fig:FMBands}
\end{center}
\end{figure}

\begin{figure}
\begin{center}
\includegraphics[width=18cm]{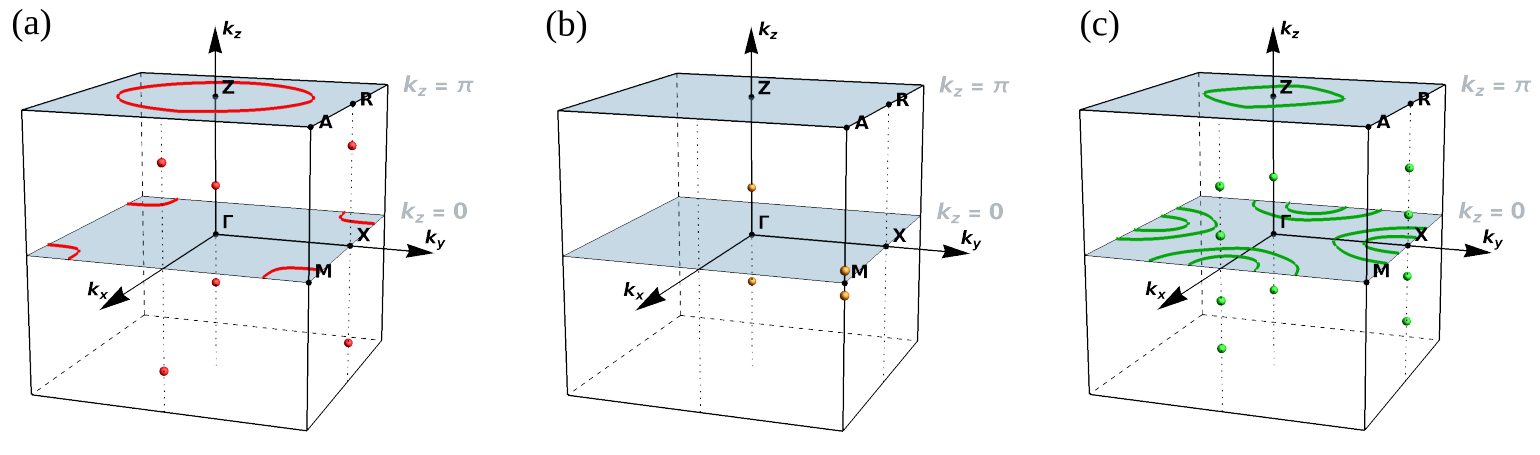}
\caption{Point and line crossings in spin-polarized SmB$_6$ with valence 2.73. (a) red, (b) orange and (c) green coloring highlights the Weyl points and nodal lines between bands (2,3), (3,4), and (4,5), respectively, see also Fig.~\ref{fig:FMBands}.
}
\label{fig:FM_Crossings}
\end{center}
\end{figure}

For sufficiently strong external magnetic fields the magnetic moments become fully polarized.
Here, we study the case when the field is along the z direction such all moments point along z.
The magnetic space group for this case is $P_4/mm'm'$ (No.~123.7.1005 in the OG convention).
This is a symmorphic space group, where only inversion, the mirror symmetry $m_z$, and rotations with respect to the z-axis remain as unitary symmetries.
We can thus expect to find accidental nodal lines in the mirror planes $k_z = 0, \pi$ and Weyl points characterized by an exchange of rotation eigenvalues along the paths $\Gamma$-Z, X-R and M-A.
In Fig.~\ref{fig:FMBands} the band structure is shown for our model, which exhibits nodal lines in the $k_z = 0$ and $k_z = \pi$ planes, and several Weyl points on all four remaining rotation axes.
An overview of all Weyl points and twofold nodal lines is given in Fig.~\ref{fig:FM_Crossings}.

\subsubsection{Topological invariants in the high-field limit}

\begin{figure}
\begin{center}
\includegraphics[width=15cm]{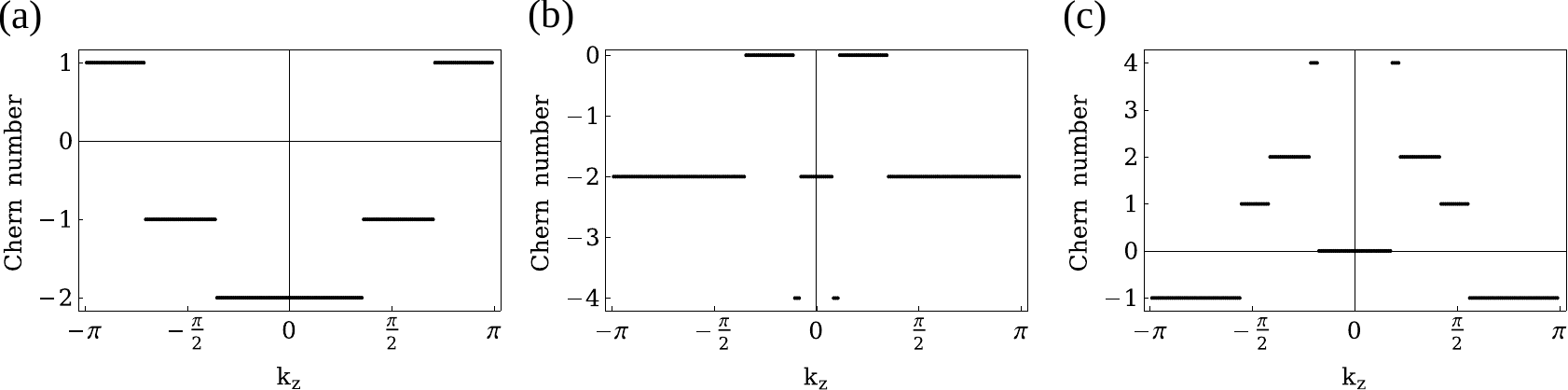}
\caption{Chern numbers for spin-polarized SmB$_6$ calculated on planes of constant $k_z$. (a),(b),(c) display the Chern numbers for fillings up to the second, third, and forth band, respectively. }
\label{fig:FM_Chernnumbers}
\end{center}
\end{figure}

For the spin-polarized case the band structure has several Weyl points of varying separation as well as nodal lines.
To characterize the Weyl points we consider their chiralities and their effect on the total Chern number calculated on planes of constant $k_z$.
In Fig.~\ref{fig:FM_Chernnumbers} we present the Chern numbers evaluated on planes perpendicular to $k_z$ for two, three and four filled bands.
Jumps of the Chern number occur only at the positions of the identified Weyl points with two exceptions.
In Fig.~\ref{fig:FM_Chernnumbers}(b) two additional jumps can be found.
Generally additional accidental Weyl points occur in a group comprising four Weyl points of identical unit charge at a given value of $k_z$ due to the crystalline symmetries.
Interestingly, the accidental Weyl points nearly coincide with a jump due to the double Weyl point along the M-A line from -2 to -4, which is followed by the change in the Chern number from -4 to 0.
Such jumps in the Chern number can only be explained by the presence of the discussed four accidental Weyl points at generic positions situated at $k_z \approx \pm 0.3$ between the bands 3 and 4.
The second case of four Weyl points at generic positions is seen in Fig.~\ref{fig:FM_Chernnumbers}(c), where the jump at $k_z \approx \pm 0.5 $  does not correspond to to any crossing on one of the rotation axes.

The Weyl points occur on both the twofold and the fourfold rotation axes and their charges are either one or two.
The latter ones (i.e. double Weyl points) are protected by the four fold rotation symmetry.
There are six Weyl points between the bands 2 and 3 with charge 1.
For the bands 3 and 4 only four double Weyl points occur, two on each of the fourfold rotation axes as well as four pairs of single Weyl points at generic positions.
Notably, the region of non-zero Chern number spans nearly the full Brillouin zone and, unlike for bands 2 and 3, the Chern number occurs effectively with a single sign, such that the contribution from different $k_z$ to the topological responses would add up.
To reach a quantitative statement regarding the topological responses, for example the anomalous Hall effect, a detailed calculation of the topological contributions by taking the partial filling of bands into account is required, but this is beyond the scope of this manuscript.
The Chern number between bands 4 and 5 takes five different values, a result of the 18 Weyl points including 8 at generic positions.

Besides the Weyl points, there are also twofold degenerate nodal lines that are protected by the symmorphic mirror symmetry $m_z$.
Since the bands are nondegenerate almost everywhere, all nodal lines exhibit a quantized Berry phase of $\theta = \pi$ on any path encircling a single line.
Notably, nodal lines occur together with some of the Weyl points in the vicinity of the Fermi level.

\begin{thebibliography}{41}
\bibitem[{note1()}]{note1}
\bibinfo{note1}{The value $\eta=0.6$ leads to an increase of the hopping parameters by 8\% at the critical pressure of $P\sim 6$~GPa, which sounds plausible.}
\end{thebibliography}
\end{document}